\documentclass{iopjournal}

\usepackage{amsmath,amsfonts,amssymb}
\usepackage{graphicx}
\usepackage{url}
\usepackage{hyperref}
\usepackage{cite}
\usepackage{algorithm}
\usepackage{algpseudocode}
\usepackage{xcolor}
\usepackage{bm}
\usepackage{comment}
\usepackage{subcaption}

\newcommand{\sign}{\mathrm{sign}}
\newcommand{\erf}{\mathrm{erf}}

\begin{document}

\articletype{Paper} 

\title{Daydreaming algorithm for Biased Patterns}

\author{Mikiya Doi$^{1,*}$\orcid{0009-0009-3305-5335}, Masayuki Ohzeki$^{1,2,3,4}$\orcid{0000-0001-9151-2914} and Federico Ricci-Tersenghi$^{5,6,7,*}$\orcid{0000-0003-4970-7376}}

\affil{$^1$Graduate School of Information Sciences, Tohoku University, Sendai 980-8579, Japan} \\
\affil{$^2$Department of Physics, Institute of Science Tokyo, Tokyo 152-8550, Japan} \\
\affil{$^3$Research and Education Institute for Semiconductors and Informatics, Kumamoto University, Kumamoto 860-8555, Japan} \\
\affil{$^4$Sigma-i Co., Ltd., Tokyo 108-0075, Japan}\\
\affil{$^5$Physics Department, Sapienza University of Rome, P.le A. Moro 5, 00185 Rome, Italy} \\
\affil{$^6$CNR--Nanotec, Rome unit, P.le A. Moro 5, 00185 Rome, Italy} \\
\affil{$^7$INFN--Sezione di Roma 1, P.le A. Moro 5, 00185 Rome, Italy}

\affil{$^*$Authors to whom any correspondence should be addressed.}

\email{mikiya.doi.t2@dc.tohoku.ac.jp, federico.ricci@uniroma1.it}

\keywords{Hopfield model, Biased patterns, Daydreaming algorithm}

\begin{abstract}
The \emph{Daydreaming} algorithm has been recently proposed in Ref.~\cite{Serricchio2025} as a learning rule that simultaneously reinforces stored patterns and suppresses spurious attractors to improve the storage capacity of the Hopfield model.
Its effectiveness has been reported for both uncorrelated and correlated data.
However, the existing formulation has mainly assumed unbiased patterns, and the formulation for biased patterns has not yet been sufficiently established.
Biased patterns are known to be much more problematic for models of associative memories.

In this study, we reformulate Daydreaming for biased patterns motivated by the underlying rationale of the pseudo-inverse rule. 
Specifically, we introduce the retrieval dynamics and an energy function based on the centered representation, and we derive a corresponding update rule for centered Daydreaming.
We compare the centered pseudo-inverse rule with centered Daydreaming for biased patterns and examine the retrieval maps and eigenvalue distributions of the coupling matrices.

As a result, the centered Daydreaming yields larger basins of attraction than the centered pseudo-inverse rule, and such a beneficial property seems to be due to the broadness of the eigenvalue spectrum of the coupling matrix.
To better understand this connection, we construct modified coupling matrices whose spectra interpolate between a pseudo-inverse-like spectrum and the Daydreaming one.
The results clearly indicate that the broader spectrum generated by Daydreaming contributes to the enlarged basin of attraction. 
\end{abstract}

\section{Introduction}

The Hopfield model is the best-known example of an attractor neural network \cite{Amari1972, Hopfield1982}.
The model embeds desired memory patterns as stable fixed points of the retrieval dynamics by appropriately designing the coupling matrix $\bm{J}$.
The original Hopfield model has neurons that take binary values $s_{i} \in \{+1, -1\}$ and evolve according to the following retrieval dynamics rule
\begin{equation}
    s_{i}^{(t+1)}
    =
    \sign \left(\sum_{j \neq i} J_{ij}s_{j}^{(t)} \right)\,.
    \label{eq:retrieval_dynamics}
\end{equation}
If the initial state lies within the basin of attraction of a stored pattern, the retrieval dynamics converge to that pattern.
Therefore, when designing the coupling matrix, not only the stability of a pattern as a fixed point but also the size of its basin of attraction are important measures.
The most classical learning rule for coupling matrices is the Hebb rule \cite{Hebb1949}.
Such a rule defines the coupling matrix $\bm{J}$ for $P$ patterns $\left\{\bm{\xi}^{\mu} \right\}_{\mu = 1, ..., P}$ with  $\xi_{i}^{\mu} \in \{+1, -1\}$, as follows
\begin{equation}
    J_{ij}
    =
    \frac{1}{N} \sum_{\mu=1}^{P} \xi_{i}^{\mu} \xi_{j}^{\mu}\;\;\text{ if } i\neq j,
    \qquad 
    J_{ii} = 0
    \label{eq:Hebb_rule}
\end{equation}
The Hebb rule has long served as the standard form of the Hopfield model because it is biologically plausible \cite{Amit1989} and tractable to theoretical analysis \cite{Amit1985}.
However, the Hebb rule has inherent limitations.
First, its storage capacity remains limited for a network with $N$ neurons, as the critical load is approximately $\alpha_{c} = P_{c}/N \approx 0.138$ \cite{Amit1985}.
Second, as the load increases, mixed states and spurious attractors other than the stored patterns become more numerous, and retrieval success depends strongly on the initial condition.
Therefore, a central challenge in the Hopfield model is to increase storage capacity and enlarge the basin of attraction while preserving biological plausibility.

Many learning rules have been proposed to address this challenge.
A representative analytical approach is the pseudo-inverse rule \cite{Personnaz1985, Kanter1987}.
This rule has a major advantage in that it can construct fixed points without retrieval errors, even for correlated patterns.
On the other hand, this approach requires the inversion of the pattern correlation matrix, which makes it non-local and not always suitable for the sequential addition of new memories.
It is also known that the basin of attraction becomes small as the load approaches the theoretical capacity.
Iterative learning rules, such as unlearning or dreaming, have also been proposed \cite{Hopfield1983}.
These methods remove undesirable stable states by progressively weakening spurious memories.
These learning rules consist of local updates, which also make them biologically plausible.
However, classical dreaming can suffer from catastrophic forgetting, in which excessive iterations deteriorate even the memories the network should retain.

Recently, a lot of research activity has explored the dreaming mechanics, and generalizations of it, with the aim of better understanding how it manages to increase the network capacity \cite{agliari2019dreaming, fachechi2019dreaming, Benedetti2022, Agliari2024, Benedetti2024, Serricchio2025, takeuchi2026analysis}.

The algorithm that at present is the most effective in increasing the capacity of the network and the size of the basin of attraction is the Daydreaming algorithm proposed in Ref.~\cite{Serricchio2025}. 
Daydreaming is a learning algorithm that maintains good performance even after long learning by combining Hebbian updates that repeatedly reinforce the patterns to be stored with anti-Hebbian (i.e., unlearning or dreaming) updates that weaken spurious attractors reached from random initial conditions.
Although Daydreaming has similarities to Boltzmann machine learning in that it combines positive and negative updates, it differs fundamentally in that the states to be unlearned are sampled via zero-temperature dynamics from random initial states, rather than by equilibrium finite-temperature sampling, so that Daydreaming is intrinsically an out-of-equilibrium process.
The authors of Ref.~\cite{Serricchio2025} showed that Daydreaming not only performs well for uncorrelated patterns, but also forms large basins of attraction for correlated patterns generated by the random features model \cite{Goldt2020, Negri2023}.
It has been further reported that Daydreaming also works effectively on more realistic data, like the MNIST dataset.
These results show that Daydreaming provides a promising approach to improve the performance of the Hopfield model.

However, the existing formulation of Daydreaming has mainly focused on unbiased patterns, where each element has zero mean, or on structured data consistent with that assumption, and a formulation for biased patterns has not yet been fully established.
We define biased patterns as a set of patterns whose elements satisfy $\mathbb{E}\left[\xi_i^{\mu} \right] = m \neq 0$.
These kinds of settings naturally arise when we consider neural representations with low activity rates \cite{schneidman2006weak}, and theory has long dealt with them as important problems \cite{Amit1987}.

Previous studies showed that a naive application of the Hopfield model to biased random patterns fails because the crosstalk noise generated by other stored patterns has a non-zero mean, which can destabilize the stored patterns even at low load \cite{Amit1987}.
They further showed that subtracting the pattern bias when constructing the coupling matrix restores the local stability of retrieval states, but is not, by itself, sufficient to ensure reliable retrieval.
Therefore, for biased patterns, the essential issue is not only how to modify the coupling matrix but also how to define dynamics consistent with the prescribed mean activity.
This issue becomes central when we extend Daydreaming for biased patterns.
A naive extension of Daydreaming is to introduce a local field that controls the bias, in addition to a centered coupling matrix obtained from subtracting the pattern means.
We believe this formulation is reasonable in itself.
However, when evaluating performance on biased patterns, learning rules such as the pseudo-inverse rule require additional hyperparameter tuning to control the mean activity, making a fair comparison difficult.

In this study, we reformulate Daydreaming for biased patterns inspired by the pseudo-inverse rule.
Specifically, we first revisit the centered pseudo-inverse rule for biased patterns and then clarify the corresponding retrieval dynamics and energy function that naturally derive from it.
Next, we introduce a Daydreaming update rule for the centered coupling matrix that is consistent with this framework.
As a result, the \emph{centered Daydreaming} we introduce in this paper is not an \textit{ad hoc} modification for biased patterns, but a learning rule defined on the same centered representation as the pseudo-inverse rule.
Moreover, the conventional Daydreaming is one of the special forms of centered Daydreaming proposed in this study.

In the remainder of this paper, we first describe the setting of biased patterns. 
We then formulate centered dynamics from the pseudo-inverse rule for biased patterns. 
On this basis, we propose the centered Daydreaming algorithm.
Finally, through numerical experiments on biased patterns, we compare centered Daydreaming with the centered pseudo-inverse rule in terms of retrieval maps, basin sizes, and spectral properties.
We further investigate how the eigenvalue spectrum affects retrieval performance using modified coupling matrices.

\section{Problem Setting and Formulation}

In this section, we first describe the setting considered in this study and then formulate centered Daydreaming for biased patterns.

\subsection{Problem Setting and Centered Representation}

We consider a Hopfield model composed of $N$ binary neurons, and we denote its state by $\bm{s} \in \{+1, -1\}^{N}$.
The stored patterns are $\left\{\bm{\xi}^{\mu} \right\}_{\mu = 1, ..., P}$ where each element $\xi_{i}^{\mu} \in \{+1, -1\}$.
In the unbiased patterns, each element has a zero mean.
In this study, we consider patterns that follow the distribution shown below, which includes unbiased patterns as a special case.
\begin{equation}
\xi_i^{\mu} =
    \left\{
    \begin{array}{ll}
        +1 & \mbox{with probability } p_1\,, \\
        -1 & \mbox{with probability } 1-p_1\,.
    \end{array}
\right.
\label{eq:biased_patterns}
\end{equation}
When $p_1 \neq 0.5$, the mean of each element is generally non-zero, and the set of patterns has a uniform bias.
We define the mean of each element across the pattern set as
\begin{equation}
    m_{i} \equiv \frac{1}{P} \sum_{\mu = 1}^{P} \xi_{i}^{\mu} \,.
    \label{eq:mean_introduction}
\end{equation}
Here, $\bm{m} = (m_1, ..., m_{N})^{T}$ denotes the neuron-wise mean.
When we deal with biased patterns, it is natural to work not with the original pattern $\bm{\xi}^{\mu}$ itself, but with a centered representation obtained by subtracting the mean.
This representation plays an important role in defining both the pseudo-inverse rule and Daydreaming consistently for biased patterns.
\begin{equation}
    \widetilde{\bm{\xi}}^{\mu} \equiv \bm{\xi}^{\mu} - \bm{m}
    \label{eq:centered_representation}
\end{equation}

\subsection{Pseudo-Inverse Rule for Biased Patterns}

For unbiased patterns, the pseudo-inverse rule defines the coupling matrix by using the pattern matrix $\bm{\Xi} = \left(\bm{\xi}^{1}, ..., \bm{\xi}^{P} \right) \in \{+1, -1\}^{N \times P}$ and the correlation matrix $\bm{C} = \frac{1}{N} \bm{\Xi}^{T} \bm{\Xi} \in \mathbb{R}^{P \times P}$ as follows
\begin{equation}
    \bm{J}
    =
    \frac{1}{N} \bm{\Xi} \bm{C}^{-1} \bm{\Xi}^{T}\,.
    \label{eq:pseudo_inverse_rule}
\end{equation}
Now, the relation $\bm{\Xi} = \bm{J} \bm{\Xi}$, together with $\xi_{i}^{\mu} \in \{+1, -1\}$, guarantees that each pattern $\bm{\xi}^{\mu}$ satisfies the fixed-point condition of the asynchronous retrieval dynamics.
For biased patterns, we use the following centered representation matrix rather than the original pattern matrix
\begin{equation}
    \widetilde{\bm{\Xi}} 
    \equiv 
    \bm{\Xi} - \bm{m}\, \bm{1}^{T}\,.
    \label{eq:centered_representation_matrix}
\end{equation}
Here, $\bm{1}$ denotes the $P$-dimensional vector whose elements are all equal to $1$.
Since the centering is performed with the empirical mean over the stored patterns, the centered representation matrix satisfies $\widetilde{\bm{\Xi}}\bm{1}=\bm{0}$.
Therefore, the corresponding centered correlation matrix is generally singular.
We denote the effective rank of the centered representation matrix by
\begin{equation}
    r 
    \equiv
    \mathrm{rank} \left(\widetilde{\bm{\Xi}}\right)
    =
    \mathrm{rank} \left(\widetilde{\bm{C}}\right).
\end{equation}
The empirical centering condition $\widetilde{\bm{\Xi}}\bm{1}=\bm{0}$ implies
$r\leq \min(N,P-1)$.
For independently generated random patterns with $0 < p_1 < 1$ and $P-1\leq N$, this centering condition is typically the only linear dependence among the columns of $\widetilde{\bm{\Xi}}$, and hence the typical rank is $r=P-1$.
We thus define the centered correlation matrix $\widetilde{\bm{C}}$ and the centered pseudo-inverse coupling matrix $\widetilde{\bm{J}}$ for biased patterns as
\begin{eqnarray}
    & \widetilde{\bm{C}} \equiv \frac{1}{N}\, \widetilde{\bm{\Xi}}^{T}\, \widetilde{\bm{\Xi}} \,,
    \label{eq:centered_correlation_matrix}
    \\
    & \widetilde{\bm{J}} \equiv \frac{1}{N}\, \widetilde{\bm{\Xi}}\, \widetilde{\bm{C}}^{+}\, \widetilde{\bm{\Xi}}^{T}\,,
    \label{eq:centered_pseudo_inverse_rule}
\end{eqnarray}
where $\widetilde{\bm{C}}^{+}$ denotes the Moore-Penrose pseudoinverse of
$\widetilde{\bm{C}}$.
Although $\widetilde{\bm{C}}^{+}\widetilde{\bm{C}}$ is a projector rather than the
identity matrix, the following relation still holds:
\begin{equation}
    \widetilde{\bm{\Xi}} = \widetilde{\bm{J}}\, \widetilde{\bm{\Xi}}\,.
    \label{eq:centered_pseudo_inverse_rule_equation1}
\end{equation}
Eq.~\eqref{eq:centered_pseudo_inverse_rule_equation1} implies that the following equation is satisfied for each pattern $\bm{\xi}^{\mu}$
\begin{equation}
    \bm{\xi}^{\mu} - \bm{m} = \widetilde{\bm{J}}(\bm{\xi}^{\mu} - \bm{m})
    \quad\Longleftrightarrow\quad
    \bm{\xi}^{\mu} = \widetilde{\bm{J}}(\bm{\xi}^{\mu} - \bm{m}) + \bm{m}\,.
    \label{eq:centered_pseudo_inverse_rule_equation2}
\end{equation}
The important point here is that $\widetilde{\bm{J}}$ contains information not on the pattern itself but on its deviation from the mean.

\subsection{Retrieval Dynamics and the Corresponding Energy Function}
Motivated by Eq.~\eqref{eq:centered_pseudo_inverse_rule_equation2}, and following the standard Hopfield convention of excluding self-couplings, we define the retrieval dynamics for biased patterns as
\begin{equation}
    s_{i}^{(t+1)}
    =
    \mathrm{sign}
    \left(\sum_{j \neq i} \widetilde{J}_{ij} \left(s_j^{(t)} - m_j \right) + m_{i} \right)\,.
    \label{eq:centered_retrieval_dynamics}
\end{equation}
This dynamics extends the retrieval dynamics for unbiased patterns to the centered representation obtained by subtracting the mean $\bm{m}$ from the state $\bm{s}$, and then considers the shift from the mean afterward.
Indeed, when $m_i = 0$ holds for all $i$, this equation reduces to the standard update rule Eq.~\eqref{eq:retrieval_dynamics}.

From the update rule in Eq.~\eqref{eq:centered_retrieval_dynamics}, we derive the corresponding energy function
\begin{equation}
    \mathcal{H}(\bm{s})
    =
    - \frac{1}{2} \sum_{i \neq j} \widetilde{J}_{ij} (s_i - m_i)(s_j - m_j)
    - \sum_{i=1}^{N} m_i s_{i}\,.
\end{equation}
The first term represents the pairwise interaction for the centered representation $\bm{s} - \bm{m}$, and the second term provides an effective field induced by the bias.

This perspective is important because it allows us to treat both the pseudo-inverse rule and Daydreaming on the same basis.
If we were to apply the pseudo-inverse rule to biased patterns in the same way as Ref.~\cite{Amit1987}, we should first construct the centered coupling matrix and then we should tune the hyperparameter $g$ that controls the mean activity, which makes a fair comparison with Daydreaming difficult.
To avoid this complication, we formulate the problem by specifying only the centered coupling matrix $\widetilde{\bm{J}}$ for biased patterns.

\subsection{Centered Daydreaming}

Daydreaming is an iterative learning rule that simultaneously reinforces the patterns to be stored and unlearns the fixed points reached from random initial states.
In Daydreaming, at each step $u$, we randomly select a stored pattern $\bm{\xi}^{\mu(u)}$, run the retrieval dynamics from a random initial state and acquire a fixed point $\bm{\sigma}^{(u)}$, and then update the coupling matrix $\bm{J}$ as follows
\begin{equation}
    J_{ij}^{(u+1)} 
    = 
    J_{ij}^{(u)} + \frac{1}{\tau N} \left(\xi^{\mu(u)}_{i} \xi^{\mu(u)}_{j} - \sigma_{i}^{(u)}\sigma_{j}^{(u)}  \right)\,.
    \label{eq:original_Daydreaming_update_rule}
\end{equation}
The parameter $\tau$ is the inverse learning rate, and can be fixed without any particular effort \cite{Serricchio2025}.

The update rule of the \emph{centered Daydreaming} we propose in this study is given as follows
\begin{equation}
    \widetilde{J}_{ij}^{(u+1)} 
    = 
    \widetilde{J}_{ij}^{(u)} + \frac{1}{\tau N} 
    \left[
        \Big(\xi^{\mu(u)}_{i} - m_i \Big) \Big(\xi^{\mu(u)}_{j} - m_j \Big)
        - 
        \Big(\sigma_{i}^{(u)} - m_i \Big) \Big(\sigma_{j}^{(u)} - m_j \Big)
    \right]\,.
    \label{eq:centered_Daydreaming_update_rule}
\end{equation}
Here, $\bm{\sigma}^{(u)}$ is the fixed point obtained by running the centered retrieval dynamics introduced in Eq.~\eqref{eq:centered_retrieval_dynamics} from a random initial state until convergence.
Algorithm \ref{alg:centered_Daydreaming} presents the pseudocode for the complete numerical procedure.

\begin{algorithm}
\caption{centered Daydreaming learning algorithm}\label{alg:centered_Daydreaming}
\begin{algorithmic}[1]
\Require patterns $\{\bm{\xi}^{\mu}\}_{\mu=1}^P$
    \State $m_{i} \gets \frac{1}{P} \sum_{\mu=1}^{P} \xi_{i}^{\mu}$ \Comment{Calculate the bias of each element}
    \State $\widetilde{J}_{ij} \gets \frac{1}{N}\sum_{\mu=1}^{P} \left(\xi_i^{\mu} - m_i \right)  \left(\xi_j^{\mu} - m_j \right)$ \Comment{Initialization with the centered Hebb rule}
    \State $\widetilde{J}_{ii} \gets 0$ 
    \For{$t=1,\dots,E$} \Comment{Do $E$ epochs}
        \For{$u=1,\dots,N$} \Comment{Do $N$ steps in each epoch}
            \State $\mu \gets \mathrm{Unif}(\{1,\dots,P\})$ \Comment{Pick a pattern at random}
            \State $\sigma_i \gets \mathrm{Unif}(\{+1, -1\}) \; \forall i$ \Comment{Initialize state at random}
            \While{not converged}
                \State $\sigma_i \gets \sign \left(\sum_{j \neq i} \widetilde{J}_{ij} (\sigma_j - m_j) + m_i \right)$ \Comment{Run the retrieval dynamics}
            \EndWhile
            \State $\widetilde{J}_{ij} \gets \widetilde{J}_{ij} + \frac{1}{\tau N}
                    \left[ 
                          (\xi_i^{\mu} - m_i) (\xi_j^{\mu} - m_j)
                          -
                          (\sigma_i - m_i) (\sigma_j - m_j)
                    \right]$ 
                    \\
                    \Comment{Update the centered coupling matrix}
            \State $\widetilde{J}_{ii} \gets 0$
        \EndFor
    \EndFor
\end{algorithmic}
\end{algorithm}

Thus, the centered Daydreaming proposed in this study is not an empirical variant obtained by just adding a bias-correction term, but a learning rule derived from the centered pseudo-inverse rule.
In the numerical experiments that follow, we apply this centered Daydreaming to biased patterns and compare it with the centered pseudo-inverse rule to show how bias affects retrieval performance, the basin of attraction, and the spectral structure of the coupling matrix.

\section{Experimental Setup}

In this section, we describe the numerical experiment setup to evaluate the performance of the centered pseudo-inverse rule and centered Daydreaming for biased patterns.

\subsection{Pattern Generation}

In this study, we use biased random patterns, where each element is generated independently, as defined in Eq.~\eqref{eq:biased_patterns}.
The case $p_1 = 0.5$ corresponds to unbiased patterns, while $p_1 > 0.5$ means biased patterns in which the positive element is dominant.
In this study, we systematically vary $p_1$ as a control parameter to examine how the strength of the bias affects retrieval performance.

\subsection{Evaluation Metrics}

We evaluate the performance of each learning rule by using the magnetization $m^{\mu}$, which is the overlap between a stored pattern $\bm{\xi}^{\mu}$ and the state $\bm{s}$
\begin{equation}
    m^{\mu} = \frac{1}{N} \sum_{i=1}^{N} \xi^{\mu}_{i} s_{i}\,.
    \label{eq:magnetization}
\end{equation}
This quantity measures how the state matches the stored pattern.
When $m^{\mu} = 1$, the state $\bm{s}$ matches the pattern $\bm{\xi}^{\mu}$ exactly.
By contrast, when $m^{\mu} = 0$, the state $\bm{s}$ is orthogonal to the pattern $\bm{\xi}^{\mu}$, which means that it contains no information about that pattern.

An important measure is the size of the basin of attraction.
Specifically, for each pattern $\bm{\xi}^{\mu}$, we prepare an initial state $\bm{s}_\text{init}$ with an initial overlap $m_\text{init}^{\mu}$ and we run the centered retrieval dynamics shown in Eq.~\eqref{eq:centered_retrieval_dynamics} until it converges to a fixed point.
We then compute the final overlap $m_\text{final}^{\mu}$ from the converged state $\bm{s}_\text{final}$.
We can interpret a learning rule as having a larger basin of attraction when it exhibits a wider plateau in $m_\text{init}^{\mu}$ such that $m_\text{final}^{\mu} \approx 1$.
Many studies have used this measure, and we also use it as a key indicator for comparison in this study.

\subsection{Experimental Details}
Throughout all numerical experiments, including the retrieval map and spectral analyses, we followed the standard Hopfield-network convention of excluding self-couplings by setting the diagonal elements of coupling matrices to zero, $\widetilde{J}_{ii} = 0$.
Since the exact projector relation of the centered pseudo-inverse matrix Eq.~\eqref{eq:centered_pseudo_inverse_rule_equation1} is defined for the matrix before removing diagonal terms, we examine the effect of retaining self-couplings in \hyperref[sec:Appendix-B]{Appendix B}.
As a representative example, we fixed $N = 500$ and $\alpha = 0.4$, varying the bias parameter $p_1$ to study the effect of the bias.
We compared the centered pseudo-inverse rule with centered Daydreaming.
For centered Daydreaming, we fixed the inverse learning rate at $\tau = 256$, a value shown to be sufficiently large for unbiased patterns in Ref.~\cite{Serricchio2025}, and examined how the performance changed over epochs.
We initialized $\widetilde{\bm{J}}$ in centered Daydreaming by using the centered Hebb rule.
To draw the retrieval map, we generated five initial states $\bm{s}_{\text{init}}$ with overlap $m_{\text{init}}^{\mu}$ for every pattern $\bm{\xi}^{\mu}$.
We performed numerical experiments on 20 instances for each value of the bias parameter $p_1$.

\section{Results}

\begin{figure}[t]
    \centering
    \includegraphics[width= 0.5\textwidth]{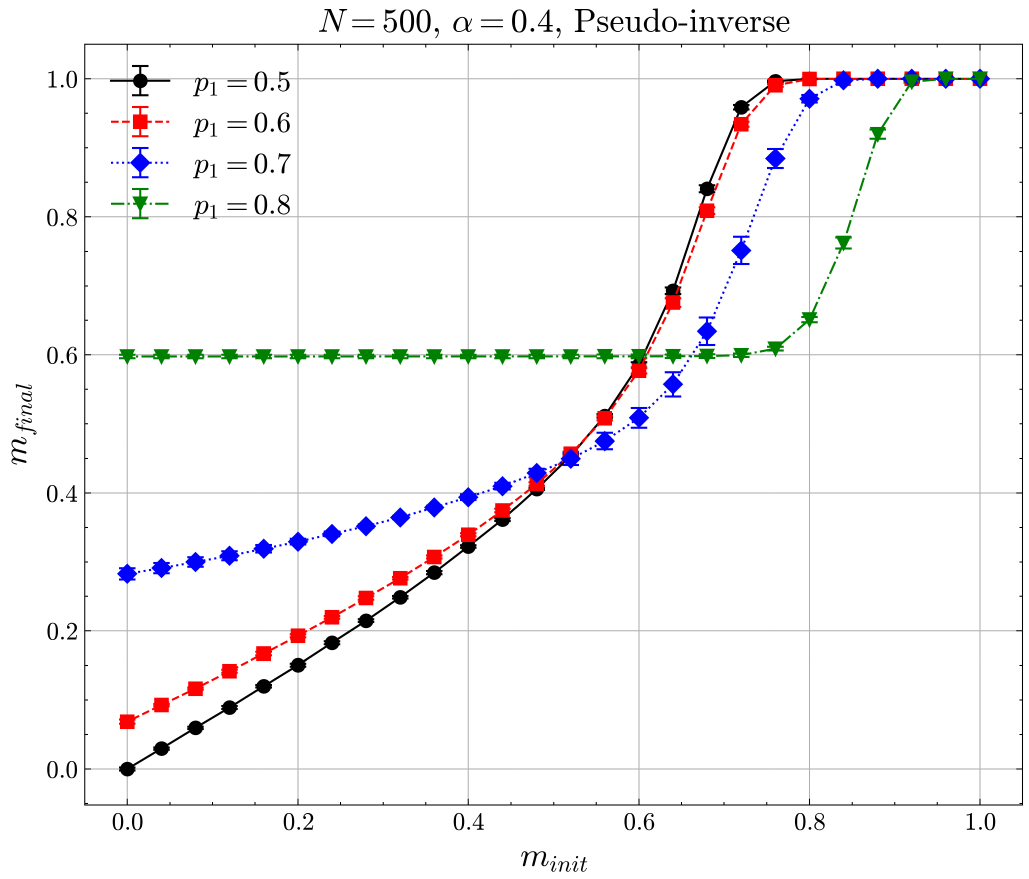}
    \caption{Retrieval maps obtained using the centered pseudo-inverse rule: each curve corresponds to a different bias strength.}
    \label{fig:retrieval_map_pseudo-inverse} 
\end{figure}

In this section, we present numerical results for the centered pseudo-inverse rule and the centered Daydreaming applied to biased patterns.
We first present the retrieval maps and the basins of attraction for each learning rule.
Next, we compare the spectral structure of the coupling matrix $\widetilde{\bm{J}}$.
Finally, we show retrieval maps of modified coupling matrices to probe how the broadness of the leading eigenvalue spectrum affects retrieval performance.

\subsection{Retrieval Maps and Basin of Attraction Size}

First, we show the retrieval maps for the centered pseudo-inverse rule in Figure~\ref{fig:retrieval_map_pseudo-inverse}.
For $p_1 = 0.5$, which corresponds to unbiased patterns, a clear plateau with $m_\text{final} \approx 1$ appears in the region with relatively large initial magnetization, $m_\text{init} \gtrsim 0.75$. 
It confirms that the pseudo-inverse rule successfully retrieves the stored patterns.
However, as we increase the bias, the shape of the retrieval map changes.
At $p_1 = 0.6$, the map still exhibits a similar plateau, but at $p_1 = 0.7$, the network requires a larger initial overlap to reach $m_\text{final} \approx 1$.
At $p_1 = 0.8$, the plateau with $m_\text{final} \approx 1$ has shrunk a lot, $m_\text{init} \gtrsim 0.9$, making pattern retrieval a very inefficient and practically impossible task even for a moderate level of noise in the data.

One may be surprised that, in this unfavorable case ($p_1=0.8$), the retrieval map in Figure~\ref{fig:retrieval_map_pseudo-inverse} shows a rather large value $m_\text{final} \approx 0.6$ at very low values of $m_\text{init}$. However, this value is what one gets if the matrix $\widetilde{\bm{J}}$ provides no information at all and centered retrieval dynamics in Eq.~\eqref{eq:centered_retrieval_dynamics} bring the system to a state where every neuron is given by $s_i=\sign(m)=1$. Indeed, in this case, one has $m_\text{final}=2p_1-1$, but this value brings no information on the pattern to be retrieved.

\begin{figure}[t]
    \centering
    \includegraphics[width=0.47\textwidth]{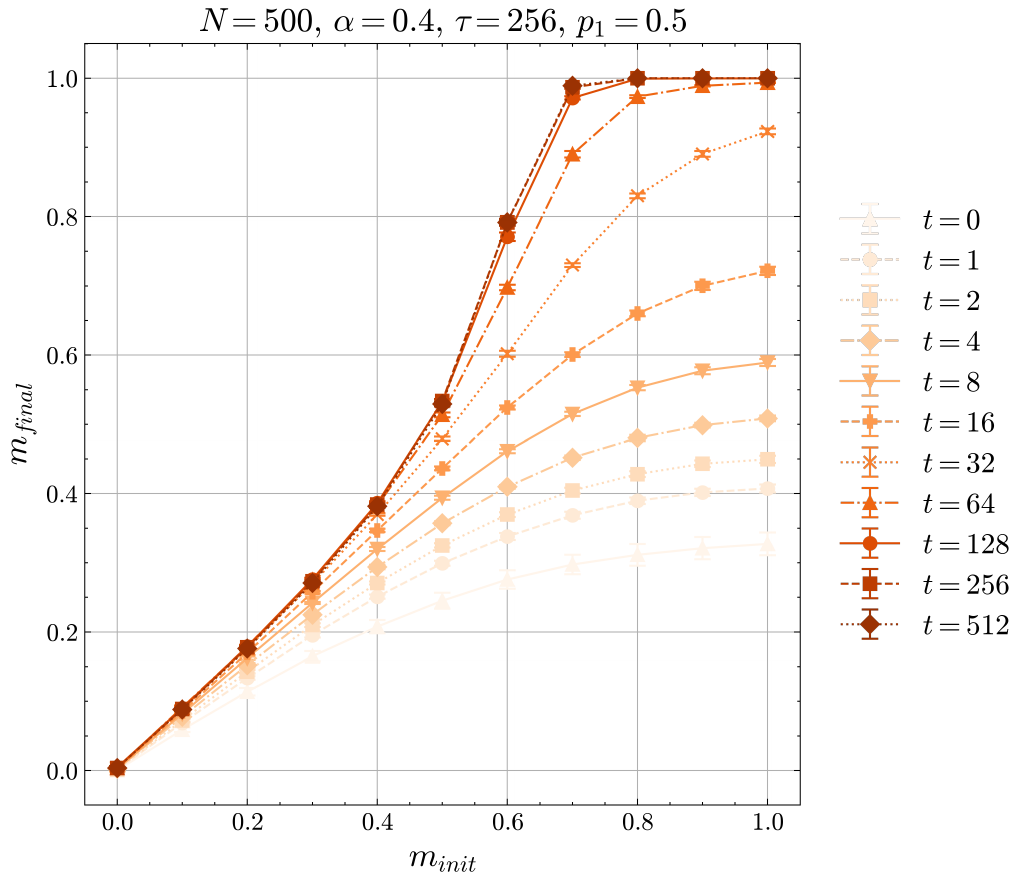}
    \hspace{5mm}
    \includegraphics[width=0.47\textwidth]{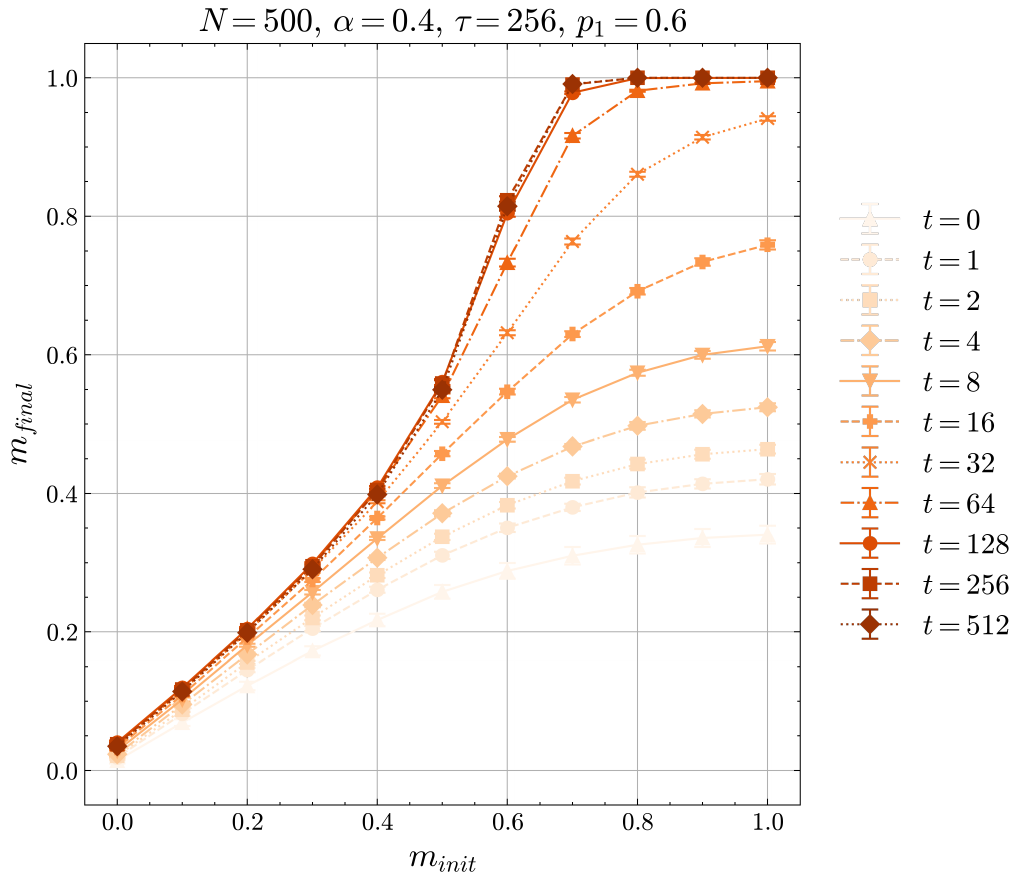}\\
    \vspace{5mm}
    \includegraphics[width=0.47\textwidth]{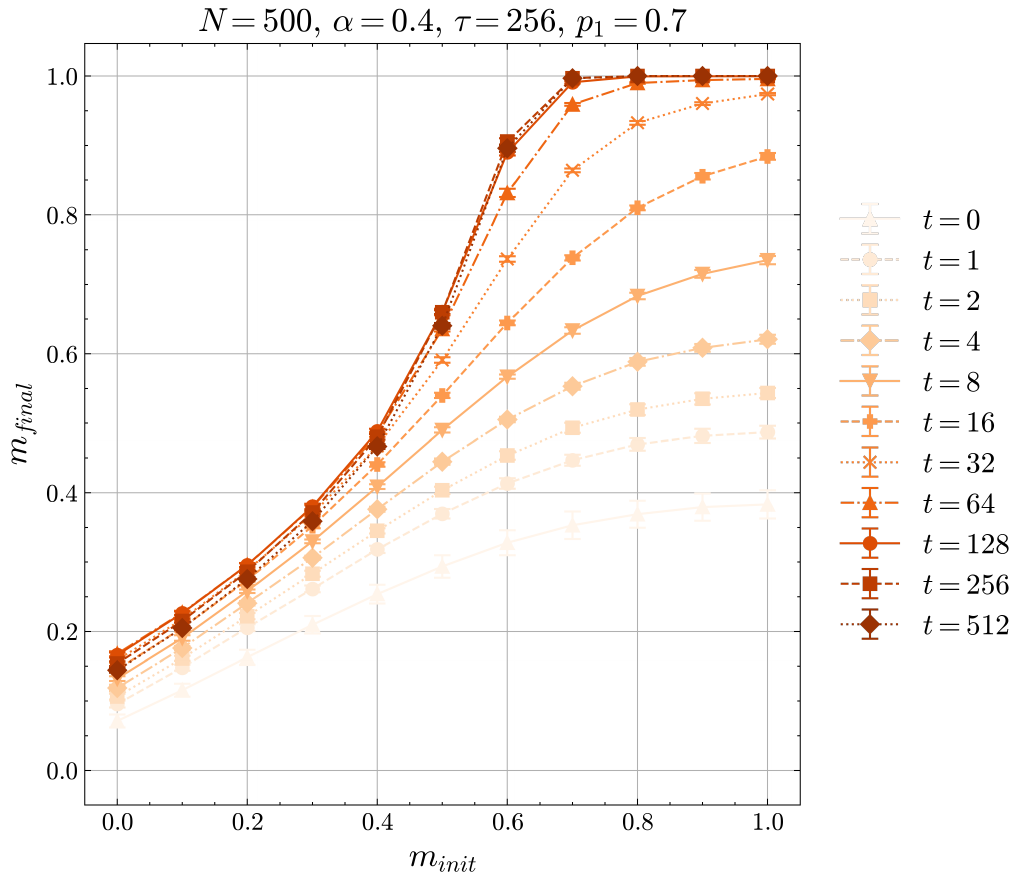}
    \hspace{5mm}
    \includegraphics[width=0.47\textwidth]{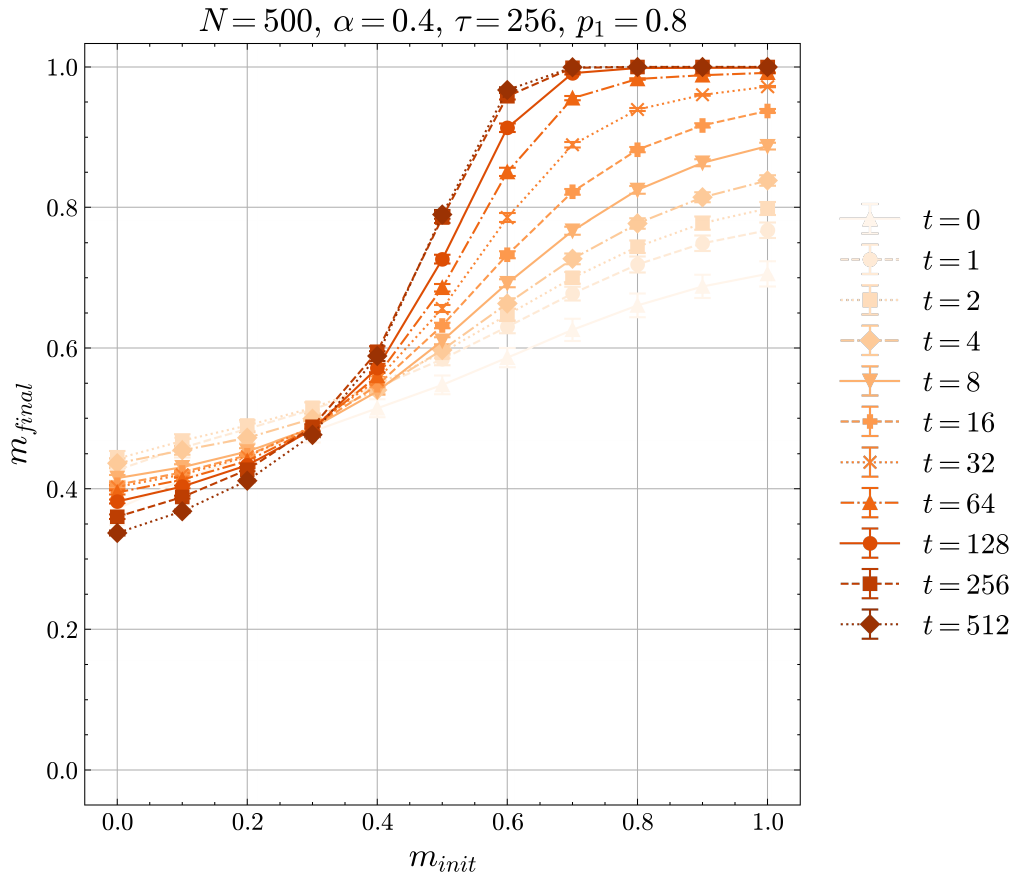}
    \caption{Retrieval maps obtained by training the coupling matrix by the centered Daydreaming algorithm: each panel differs in the value of bias parameter $p_1$.}
    \label{fig:retrieval_map_centered_Daydreaming}
\end{figure}

Next, we present the retrieval maps for the centered Daydreaming in Figure \ref{fig:retrieval_map_centered_Daydreaming}.
At $p_1 = 0.5$ (upper left panel), the retrieval map improves monotonically with the number of learning epochs, and after a sufficiently long learning, a plateau with $m_\text{final} \approx 1$ emerges for $m_\text{init} \gtrsim 0.7$, in agreement with the results of Ref.~\cite{Serricchio2025}.
The key observation is that the same plateau ($m_\text{final} \approx 1$ for $m_\text{init} \gtrsim 0.7$) is reached after learning by centered Daydreaming for any bias strength ($0.5\le p_1 \le 0.8$) that we have studied.
This means the centered Daydreaming algorithm is robust with respect to the bias in the data to be stored in a Hopfield model.
This finding makes the centered Daydreaming algorithm a general and robust procedure to train a Hopfield network with realistic data.
Notice that the centered Daydreaming algorithm has just one parameter, $\tau$, which can be fixed without any particular effort (as in the original Daydreaming algorithm).
So, centered Daydreaming is a ready-to-use, general and robust algorithm.

We also checked the dependence on the load, $\alpha = P/N$.
The results reported in \hyperref[sec:Appendix-C]{Appendix C} show that centered Daydreaming retains plateaus even when $\alpha$ is increased, both for unbiased and biased patterns.
This confirms that the robustness to bias observed at $\alpha = 0.4$ is not restricted to a single load value.

\subsection{Spectral Properties of the Coupling Matrix}

\begin{figure}[t]
    \centering
    \includegraphics[width=0.47\textwidth]{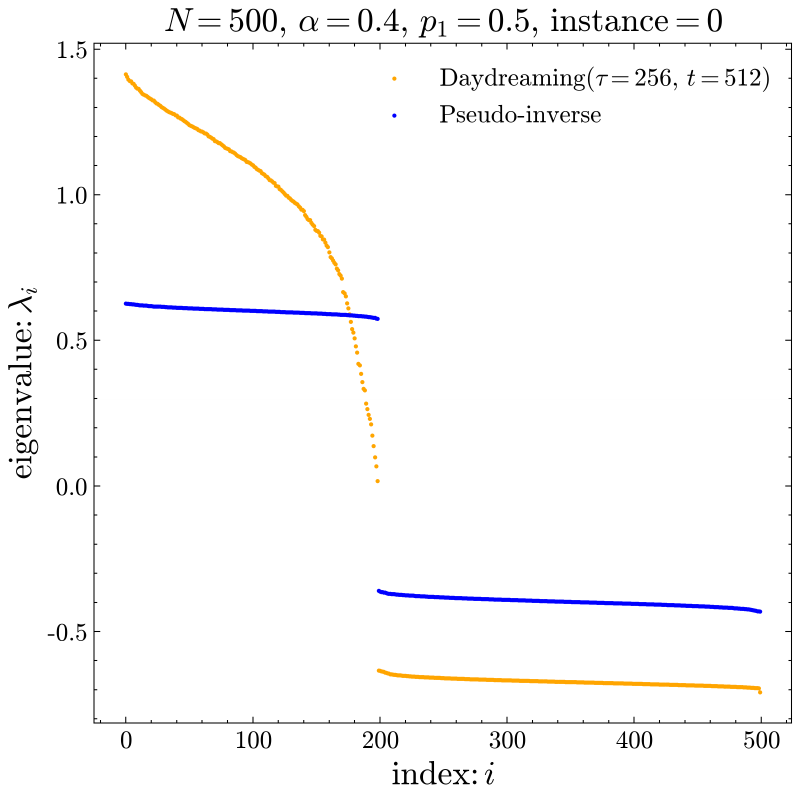}
    \hspace{5mm}
    \includegraphics[width=0.47\textwidth]{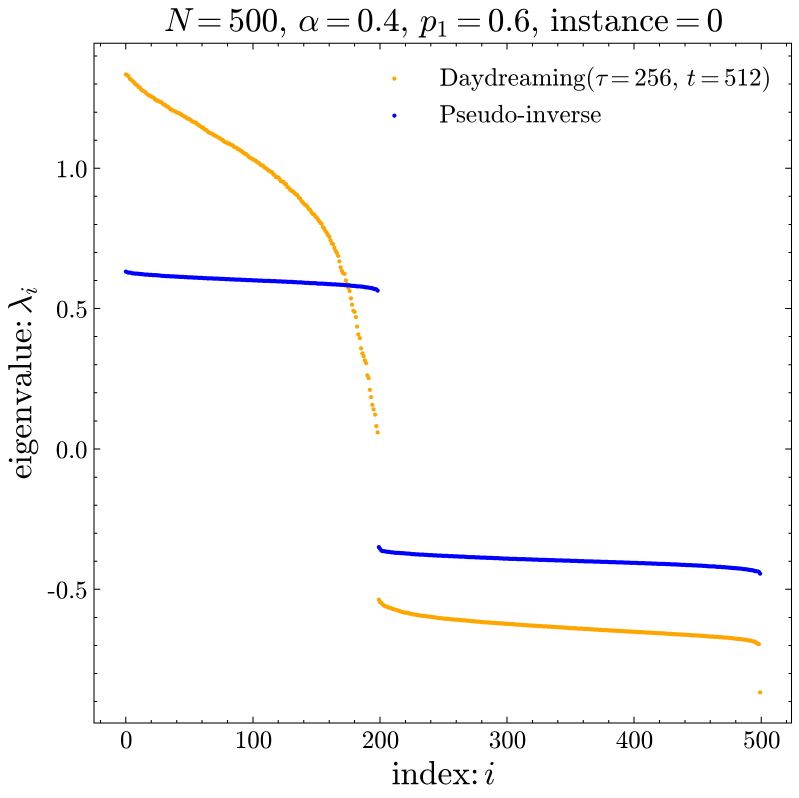}\\
    \vspace{5mm}
    \includegraphics[width=0.47\textwidth]{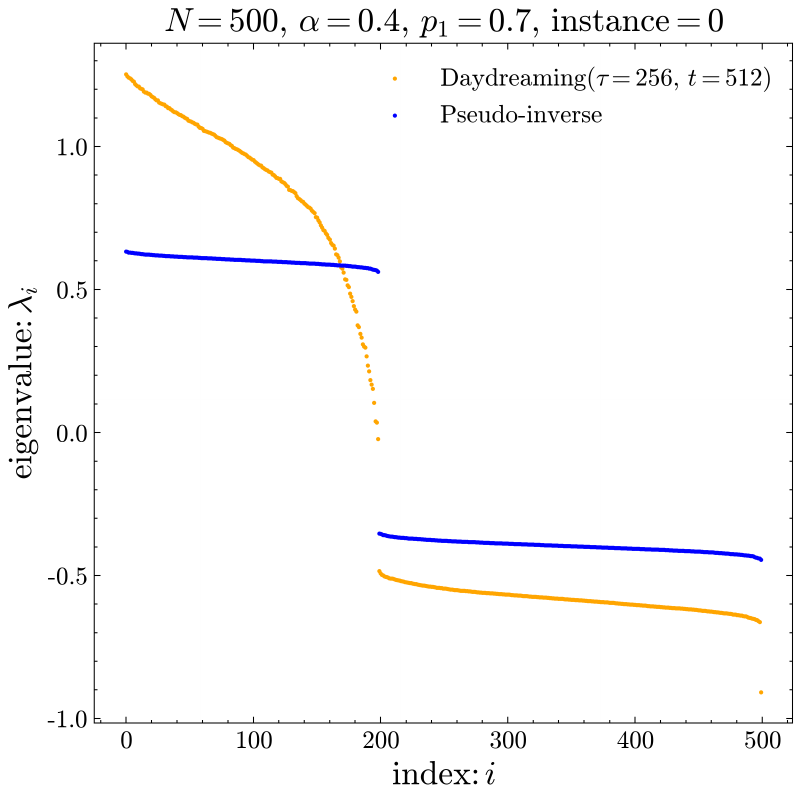}
    \hspace{5mm}
    \includegraphics[width=0.47\textwidth]{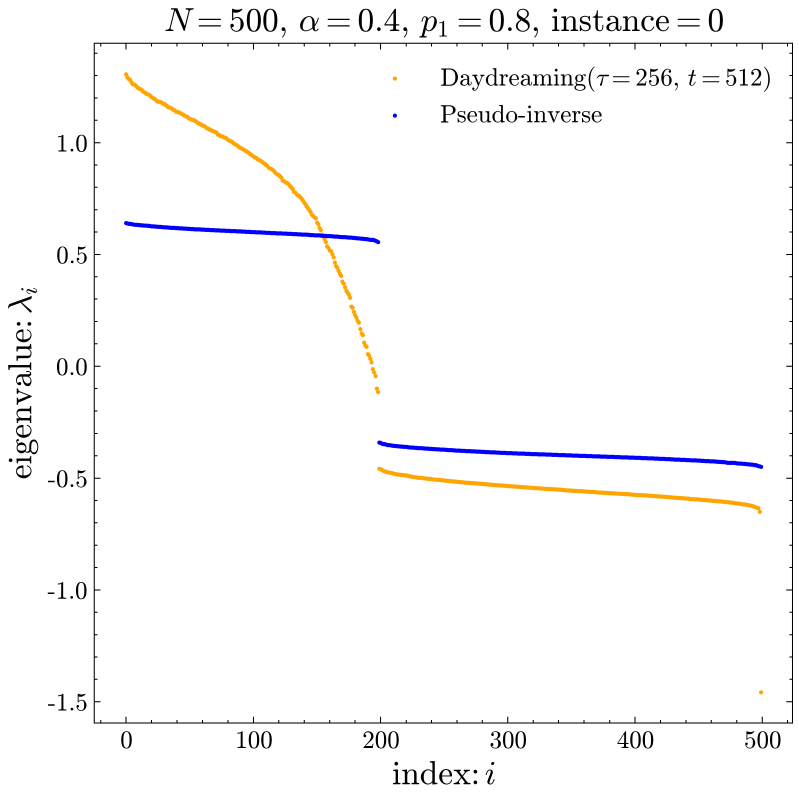}
    \caption{Eigenvalue distributions for the pseudo-inverse rule and Daydreaming. For Daydreaming, we plot the eigenvalue distribution at epoch $t=512$ for a representative instance.}
    \label{fig:eigenvalue_distribution}
\end{figure}

The eigenvalue distributions of $\widetilde{\bm{J}}$ obtained from the centered pseudo-inverse rule and the centered Daydreaming differ for every value of $p_1$ as shown in Figure~\ref{fig:eigenvalue_distribution}.

The spectrum of the centered pseudo-inverse rule is concentrated in two spectral sectors: the leading $r=P-1$ modes are associated with the subspace spanned by the centered representations of the $P$ patterns and are responsible for bringing the retrieval dynamics towards one of the patterns; the remaining $N-r$ modes correspond to the null space (recall that the coupling matrix has rank $r=P-1$ because of the centering condition).
The double-peak shape of the spectrum is preserved for all $p_1$ values.

The spectrum of the matrix learned via the centered Daydreaming algorithm is different: the negative peak (corresponding to the null space) is still very narrow, but the distribution of positive eigenvalues is much broader, with a maximum eigenvalue that is more than twice that of the pseudo-inverse and a tail extending down to the origin.
What we observe in Figure~\ref{fig:eigenvalue_distribution} for Daydreaming is consistent with the picture reported in previous studies on dreaming-based learning, where the learning process selectively reshapes the upper part of the spectrum of the coupling matrix \cite{Benedetti2024}.

Given that the coupling matrix obtained by Daydreaming shows better statistical properties in terms of basins of attraction and, consequently, in terms of maximum capacity, it is worth trying to relate this better performance to the broadness of the spectrum shown in Figure~\ref{fig:eigenvalue_distribution}.
In both cases, pseudo-inverse and Daydreaming, the leading $r$ eigenvectors are expected to span the subspace defined by the centered representations of the $P$ patterns.
Thus, the application of the matrix $\widetilde{\bm{J}}$ to the centered state vector $\bm{s}-\bm{m}$ during the retrieval dynamics in Eq.~\eqref{eq:centered_retrieval_dynamics} tends to project the centered state into this subspace.
What differs between the two rules is that, in the pseudo-inverse case, all these leading modes have approximately the same weight, while in the Daydreaming case, the weights used to project the centered state vector are very diverse.
This induces a sort of hierarchy among eigenvectors in the Daydreaming case, such that the centered state vector is first projected towards the eigenvectors with the largest eigenvalues, almost ignoring the other eigenvectors; these are considered in later stages of the dynamics.
This observation suggests that the enlarged basin of attraction is related not only to the subspace spanned by the leading $r$ eigenvectors, but also to the non-uniform weights assigned to these modes by the first $r$ eigenvalues.
To evaluate this point quantitatively, in the next subsection, we construct modified coupling matrices that keep the eigenvectors fixed while controlling the broadness of the leading $r$ eigenvalues.

\subsection{Retrieval Maps of Modified Coupling Matrices}
We probe how the broadness of the leading $r=P-1$ modes discussed in the previous subsection affects retrieval performance. 
Specifically, we construct modified coupling matrices that interpolate between the broad leading spectrum generated by Daydreaming and a pseudo-inverse-like spectrum.
We then use these modified coupling matrices to draw retrieval maps and quantify the relationship between spectral broadness and retrieval performance.

\begin{figure}
    \centering
    \includegraphics[width=0.47\linewidth]{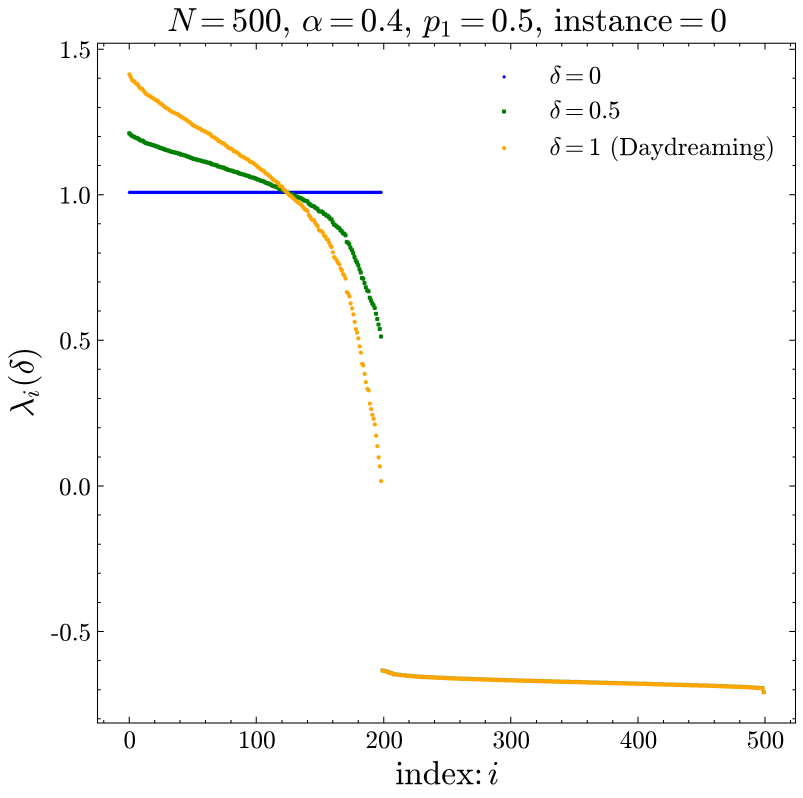}
    \hspace{5mm}
    \includegraphics[width=0.47\linewidth]{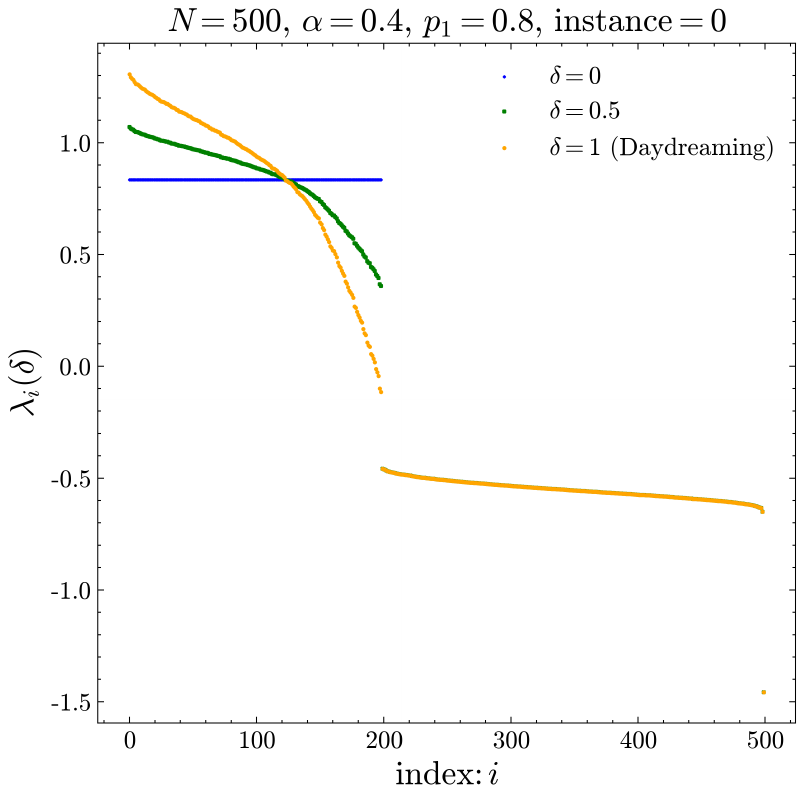}
    \caption{Designed eigenvalue spectra parametrized by $\delta$ and used to construct the modified coupling matrices, shown for a representative instance: each panel differs in the value of bias parameter $p_1$.}
    \label{fig:modified_eigenvalue_distribution}
\end{figure}

First, we perform a spectral decomposition of the coupling matrix learned by Daydreaming and obtain its eigenvalues and eigenvectors. 
We use the Daydreaming coupling matrix at $t=512$ as the original coupling matrix to be spectrally decomposed. 
We denote this matrix by $\widetilde{\bm{J}}_{\text{Daydreaming}}$ and write its spectral decomposition as
\begin{equation}
    \widetilde{\bm{J}}_{\text{Daydreaming}}
    =
    \sum_{i=1}^{N}
    \lambda_i \bm{v}_i \bm{v}_i^{T},
    \qquad
    \lambda_1 \geq \lambda_2 \geq \cdots \geq \lambda_N .
    \label{eq:spectral_decomposition}
\end{equation}
Here, we assume that the eigenvalues are sorted in descending order. 
Next, for the leading $r$ eigenvalues associated with the subspace spanned by the centered representations of the $P$ patterns, we introduce a transformation parameterized by $\delta\in[0,1]$. Defining the mean value $\bar{\lambda}=\sum_{\mu=1}^{r}\lambda_{\mu}/r$, we have
\begin{align}
    \lambda_{\mu}(\delta) =
    (1-\delta)\bar{\lambda}
    +
    \delta \lambda_{\mu},
    \qquad
    \mu=1,\ldots, r .
    \label{eq:modified_leading_eigenvalues}
\end{align}
The case $\delta=1$ corresponds to the original Daydreaming spectrum, whereas $\delta=0$ corresponds to a pseudo-inverse-like setting in which all leading $r$ modes have equal weights.
The remaining $N-r$ eigenvalues and all eigenvectors are kept unchanged.
We show specific examples of modified spectra for $p_1 = 0.5$ and $p_1 = 0.8$ in Figure \ref{fig:modified_eigenvalue_distribution}.
Using the spectrum characterized by $\delta$, we construct the modified coupling matrix as
\begin{equation}
    \widetilde{\bm{J}}_{\text{modified}}(\delta)
    =
    \sum_{\mu=1}^{r}
    \lambda_{\mu}(\delta)
    \bm{v}_{\mu} \bm{v}_{\mu}^{T}
    +
    \sum_{\nu=r+1}^{N}
    \lambda_{\nu}
    \bm{v}_{\nu} \bm{v}_{\nu}^{T}.
    \label{eq:modified_coupling_matrix}
\end{equation}

\begin{figure}
    \centering
    \includegraphics[width=0.47\linewidth]{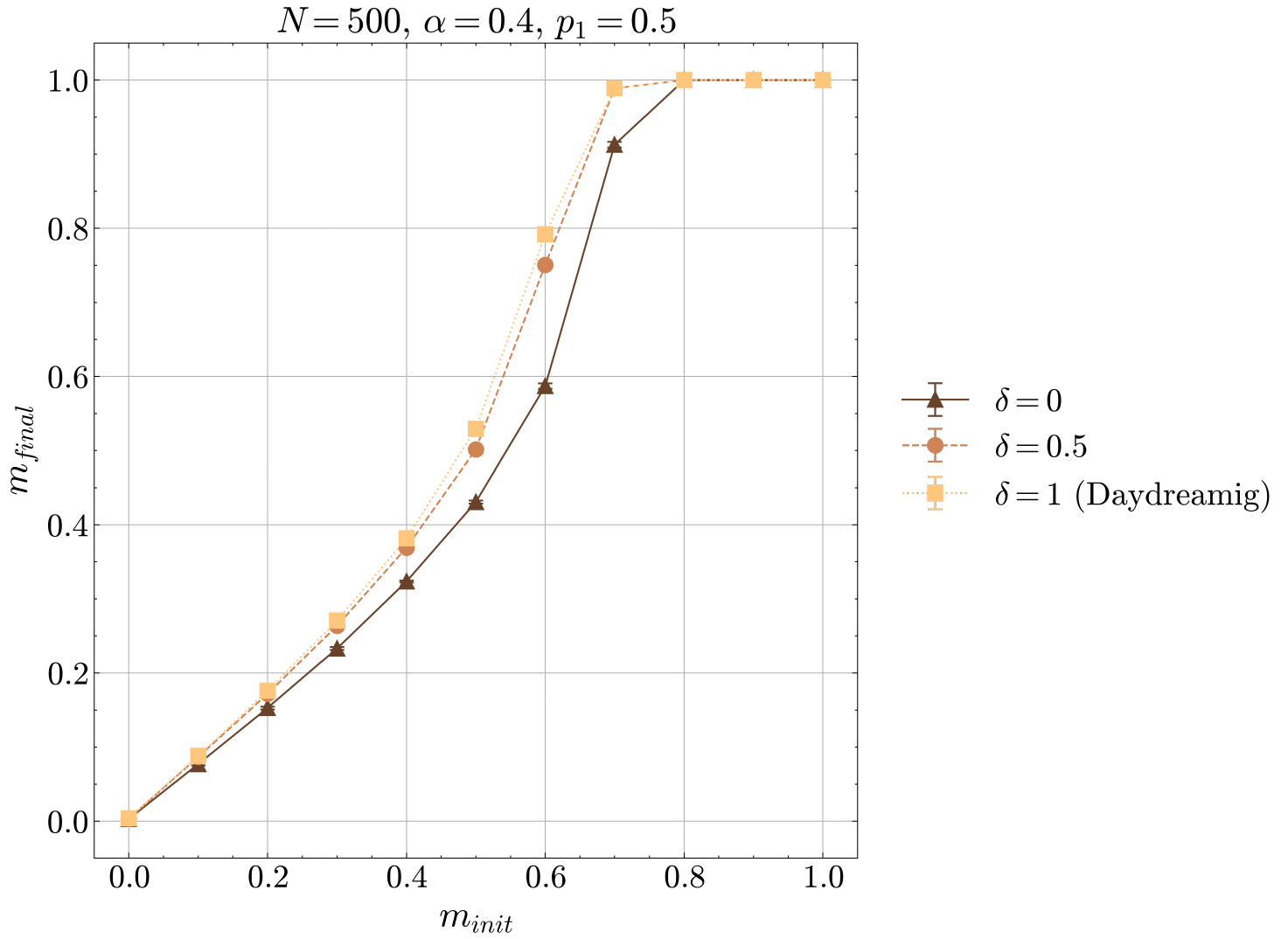}
    \hspace{5mm}
    \includegraphics[width=0.47\linewidth]{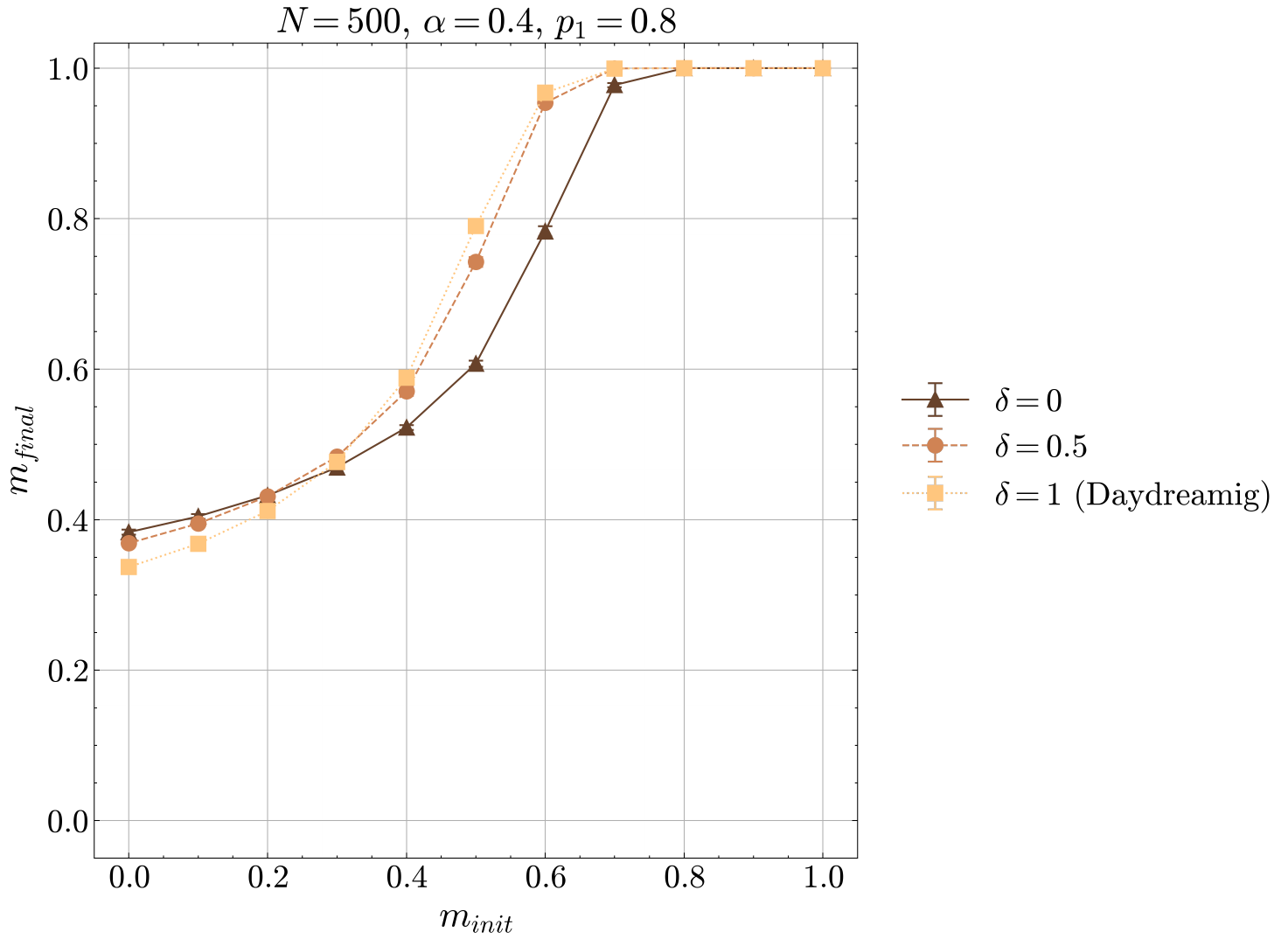}
    \caption{Retrieval maps of the modified coupling matrices for $p_1=0.5$ and $p_1=0.8$. 
The parameter $\delta$ controls the broadness of the leading eigenvalue spectrum, with $\delta=1$ corresponding to the original Daydreaming spectrum and $\delta=0$ to the pseudo-inverse-like spectrum.}
    \label{fig:retrieval_map_modified_coupling_matrix}
\end{figure}

We examine the retrieval maps obtained by varying $\delta$ in the modified coupling matrix~\footnote{Before using the modified coupling matrices $\tilde{\bm{J}}_{\text{modified}}(\delta)$ in the retrieval dynamics, we apply the same zero-diagonal convention as in the other experiments.}.
For each of $p_1 = 0.5$ and $p_1 = 0.8$, we construct the modified coupling matrices for all 20 instances and draw the corresponding retrieval maps. 
We show the results in Figure \ref{fig:retrieval_map_modified_coupling_matrix}. 
As a result, for both $p_1 = 0.5$ and $p_1 = 0.8$, we observed that the plateau region with $m_{\text{final}} \approx 1$ remains unchanged going from $\delta=1$ to $\delta=0.5$, while it clearly shrinks at $\delta = 0$. 
Thus, we conclude that the spectral broadness (keeping eigenvectors fixed) certainly contributes to the enlargement of the basins of attraction in the retrieval dynamics.

\section{Conclusions}

In this study, we have introduced a \emph{centered} version of the Daydreaming algorithm that can work effectively in the tested biased settings.
Increasing the bias in the data, the coupling matrix trained by the centered Daydreaming algorithm presents larger basins of attraction with respect to the centered pseudo-inverse and to the original Daydreaming algorithm.
Training by centered Daydreaming, the size of the basin of attraction stays almost constant in a broad range of biases ($0.5 \le p_1 \le 0.8$).
These results make the centered Daydreaming algorithm of practical use in realistic situations where data can be structured and biased.

We have also shown that the coupling matrix learned by centered Daydreaming has a broader eigenvalue spectrum than that obtained with the centered pseudo-inverse rule.
Experiments with modified coupling matrices, in which the leading $r$ eigenvalues were averaged out toward their mean while the eigenvectors were kept fixed, provide evidence that the broadness of the weights/eigenvalues of the leading modes is associated with enlarged basins of attraction.
They show that a pseudo-inverse-like leading spectrum shrinks the retrieval plateau, indicating that the non-uniform weighting of the leading $r$ modes affects retrieval performance and contributes to enlarging the basin of attraction.
This finding suggests that centered Daydreaming shapes a richer, possibly hierarchical, energy landscape that helps retrieval dynamics.

Supplementary experiments (shown in the Appendices) further support the effectiveness of centered Daydreaming. 
The load-dependence shows that the robustness to bias is not restricted to the representative load.
The biased Random-Features Hopfield Model experiment provides an additional example in which centered Daydreaming forms retrieval plateaus for structured biased patterns. 
However, the slight degradation after long learning epochs and the behavior of feature magnetization indicate that the mechanism in correlated biased data is not yet fully understood.

Future work will include a more precise analysis of how the breadth of the leading eigenvalue spectrum affects retrieval performance, and a deeper understanding of centered Daydreaming for more structured, biased patterns, such as biased Random-Features Hopfield Model across a wider range of $\alpha$ and $\alpha_D$.

\section*{Appendix A. Original Daydreaming for Biased Patterns}

\begin{figure}[t]
    \centering
    \includegraphics[width=0.47\textwidth]{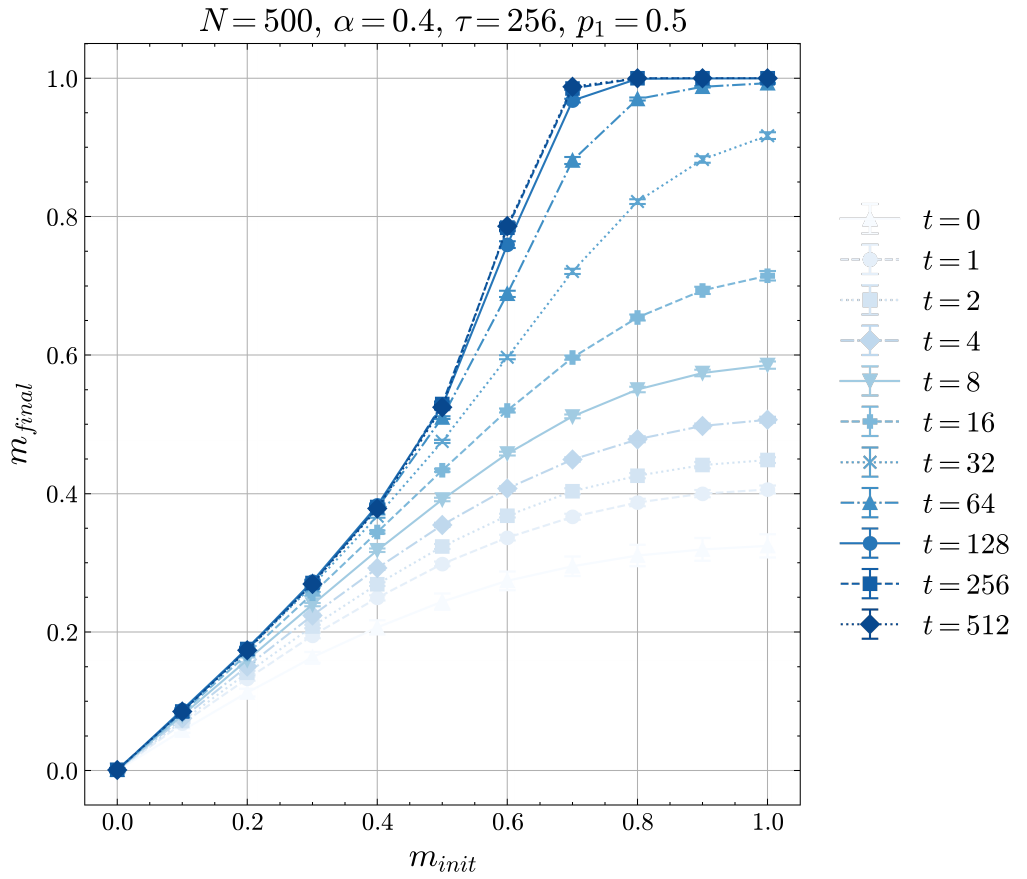}
    \hspace{5mm}
    \includegraphics[width=0.47\textwidth]{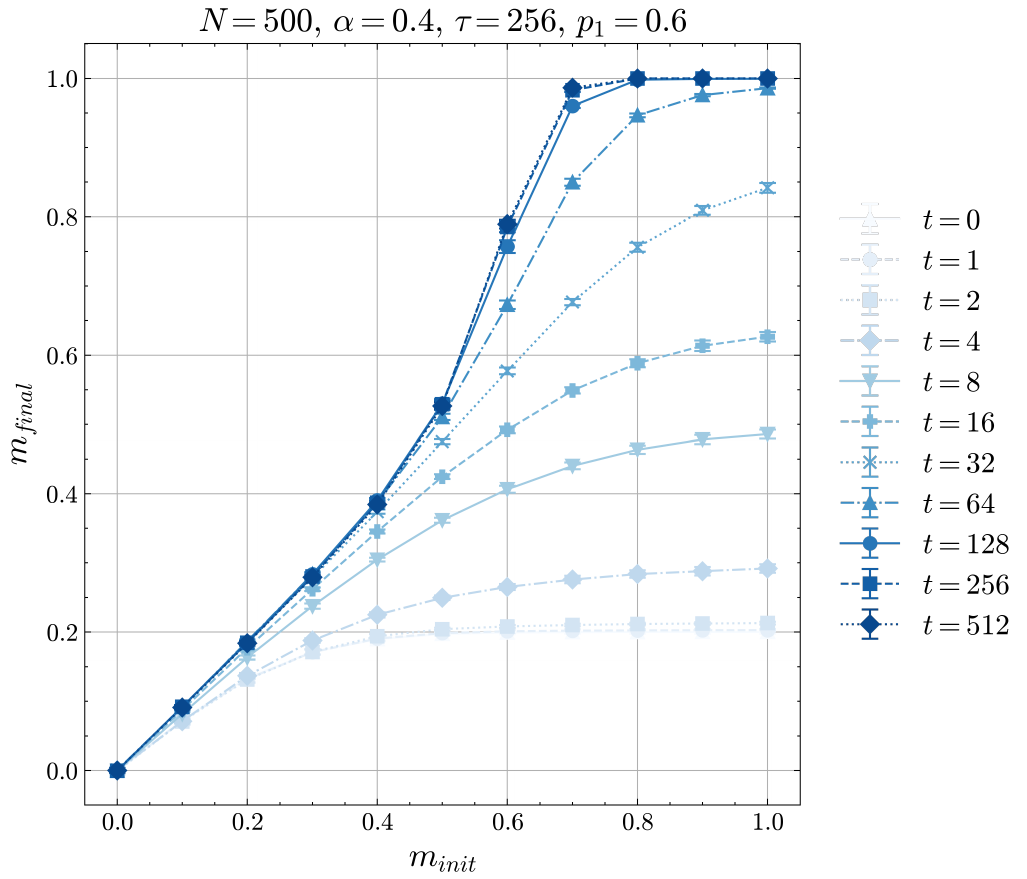}\\
    \vspace{5mm}
    \includegraphics[width=0.47\textwidth]{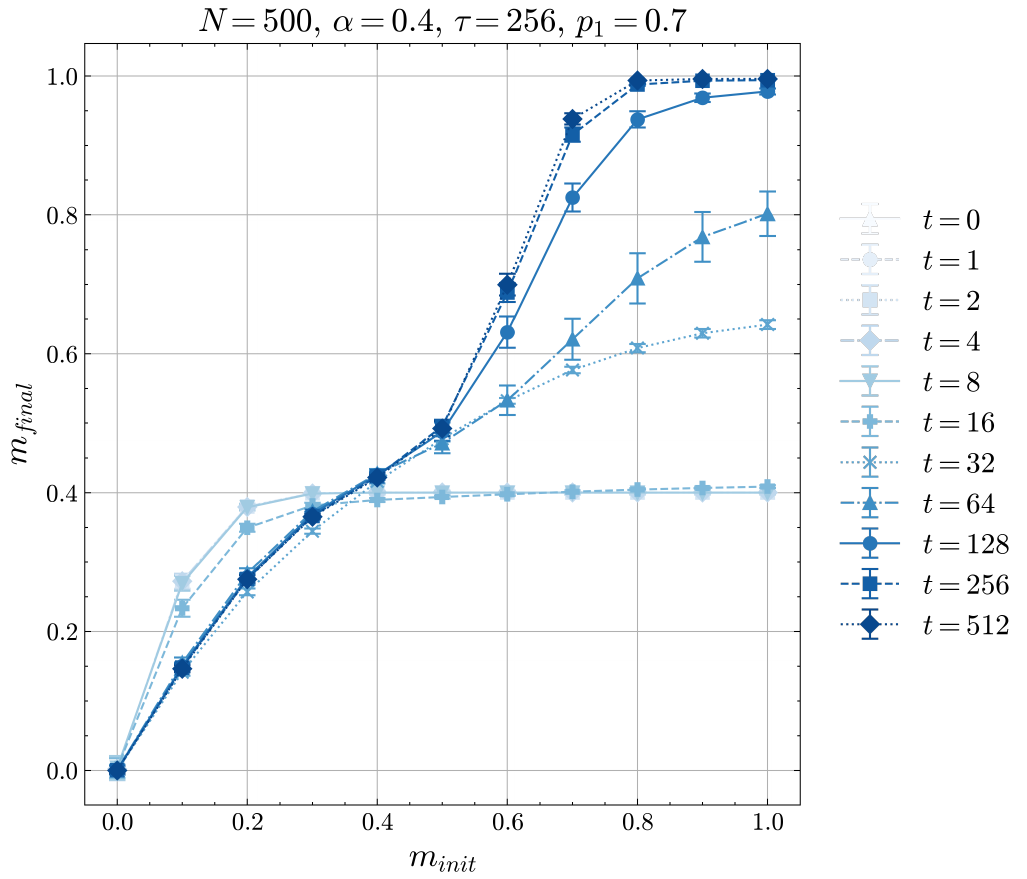}
    \hspace{5mm}
    \includegraphics[width=0.47\textwidth]{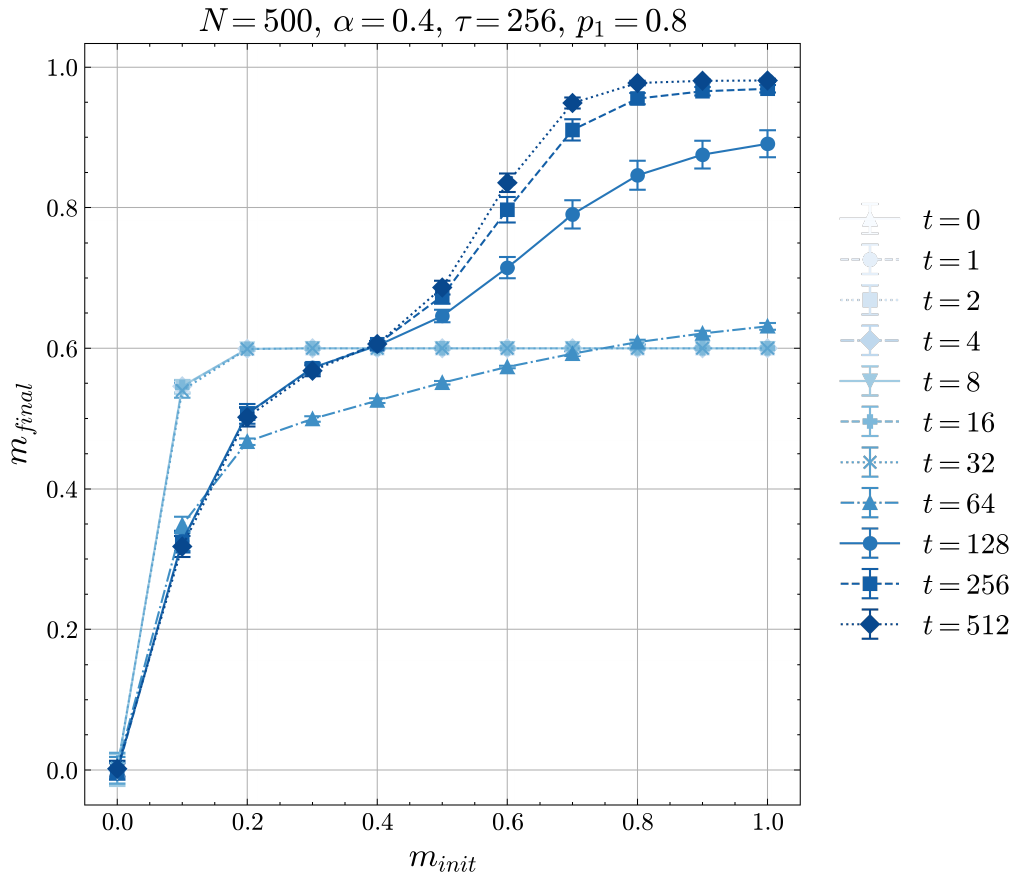}
    \caption{Retrieval map for the original Daydreaming algorithm: panels differ by the value of bias parameter $p_1$.}
    \label{fig:retrieval_map_original_Daydreaming}
\end{figure}

For completeness, we also present in Figure~\ref{fig:retrieval_map_original_Daydreaming} the results obtained by applying the original Daydreaming algorithm of Ref.~\cite{Serricchio2025} with biased patterns.
With respect to the results obtained by the centered version and reported in Figure~\ref{fig:retrieval_map_centered_Daydreaming}, when the bias parameter $p_1$ becomes large, we observe that (i) the plateau with $m_\text{final} \approx 1$ is smaller and (ii) the time required by the learning process to converge is larger. For example, in the $p_1=0.8$ case (lower right panel), at time $t=512$ the learning process has still not reached convergence, and indeed the plateau with $m_\text{final} \approx 1$ is still not formed.
Clearly, the centered version of the Daydreaming algorithm introduced in the present work has much better performance than the original Daydreaming algorithm.

\section*{Appendix B. Effect of self-couplings in the centered pseudo-inverse rule}
\label{sec:Appendix-B}
\begin{figure}[t]
    \centering
    \includegraphics[width= 0.5\textwidth]{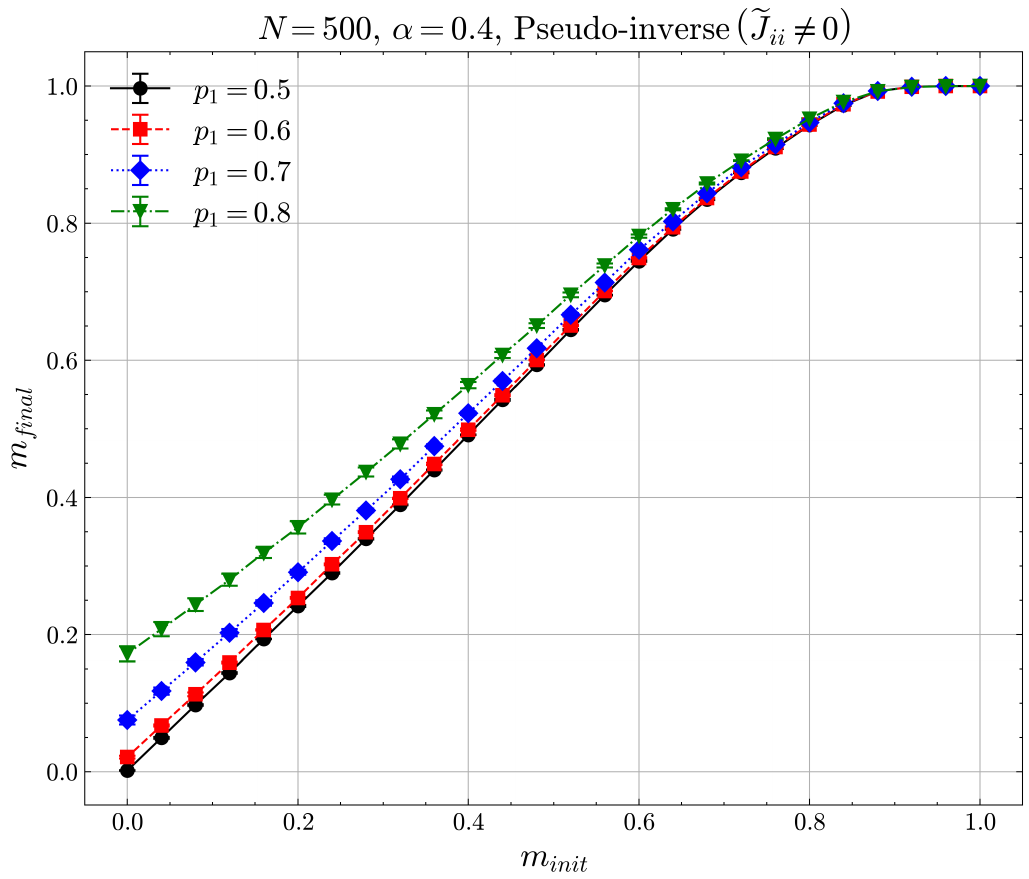}
    \caption{Retrieval maps obtained using the centered pseudo-inverse rule with self-couplings $\widetilde{J}_{ii}$: each curve corresponds to a different bias strength.}
    \label{fig:self_coupled_retrieval_map_pseudo-inverse} 
\end{figure}

In the main text, we evaluated the retrieval performance of the centered pseudo-inverse rule, excluding the effect of self-couplings.
However, the exact projector relation of the centered pseudo-inverse matrix is defined for the matrix before the diagonal interactions are removed.
Therefore, to clarify the role of the diagonal interactions, we also evaluate the centered pseudo-inverse dynamics with self-couplings included:
\begin{equation}
    s_{i}^{(t+1)}=
    \sign
    \left(\sum_{j=1}^{N} \widetilde{J}_{ij} \left(s_j^{(t)} - m_j \right) + m_i \right).
    \label{eq:self_coupled_centered_retrieval_dynamics}
\end{equation}
This dynamics preserves the exact fixed-point property of the full centered pseudo-inverse matrix.
We show the retrieval map at $\alpha=0.4$ in Figure~\ref{fig:self_coupled_retrieval_map_pseudo-inverse}.
Despite the exact fixed-point property, retaining self-couplings does not improve the retrieval performance drastically.
This result indicates that exact stability of the stored patterns does not necessarily imply a large basin of attraction.
We can understand this behavior from the form of the self-coupled local field.
Including self-couplings, the retrieval dynamics differ from the one in Eq.~\eqref{eq:centered_retrieval_dynamics} by the additional term $\widetilde{J}_{ii} \left(s_i^{(t)}-m_i\right)$.
This term depends on the current neuron value and therefore acts as an inertia-like contribution because typically $\widetilde{J}_{ii} > 0$.
It stabilizes neurons that are already aligned with a stored pattern, but it can also hold erroneous neurons in noisy initial states.
Consequently, restoring the exact fixed-point property by retaining self-couplings does not necessarily enhance the dynamics' error-correction ability.
These results suggest that the superiority of centered Daydreaming over the centered pseudo-inverse rule cannot be attributed simply to the removal of self-couplings from the centered pseudo-inverse matrix.

\section*{Appendix C. Load dependence with respect to $\alpha$}
\label{sec:Appendix-C}
In the main text, we fixed the load at $\alpha=P/N=0.4$ to clearly compare the effect of the bias parameter $p_1$.
In this Appendix, we present additional experiments in which $\alpha$ is varied to examine whether the robustness of centered Daydreaming against bias observed in the main text is specific to this particular load value.

The experimental setup is the same as in the main text. 
We set $N=500$ and used $P=\alpha N$ stored patterns. 
For each set of patterns, we computed the neuron-wise mean $m_i$ and learned the coupling matrix using centered Daydreaming. 
We fixed the inverse learning rate to $\tau=256$, as in the main experiments.
We considered $\alpha = \{0.05, 0.2, 0.6, 1.0\}$, and examined both the unbiased case with $p_1=0.5$ and the strongly biased case with $p_1=0.8$. 
We set the number of learning epochs to $t=512$ for $\alpha=0.05, 0.2,$ and $0.6$. 
For $\alpha=1.0$, we needed a longer learning time $t=8192$ because the convergence of learning becomes slower.
We performed numerical experiments on 20 instances for each setting.

\begin{figure}
    \centering
    \includegraphics[width=0.49\linewidth]{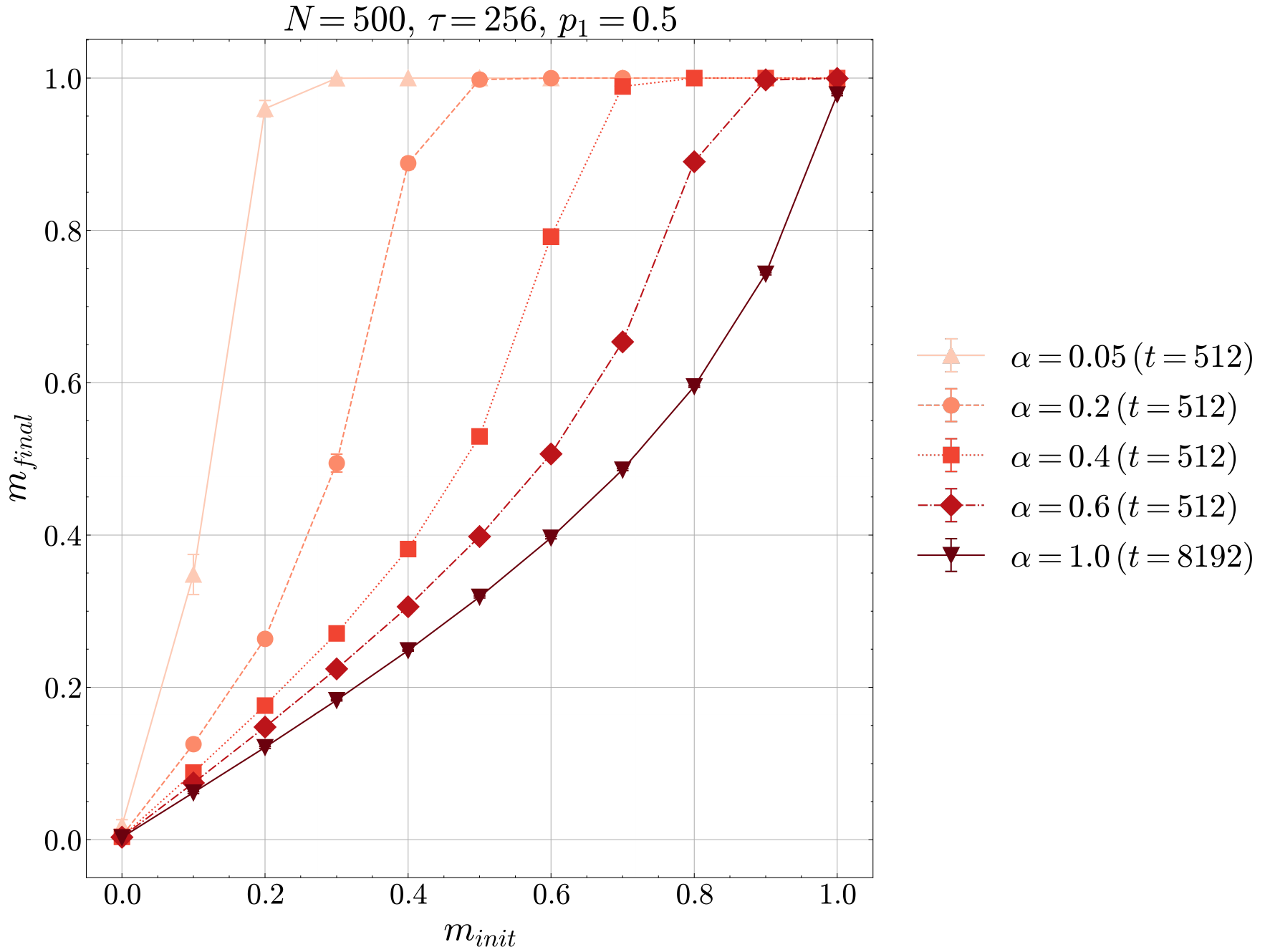}
    \includegraphics[width=0.49\textwidth]{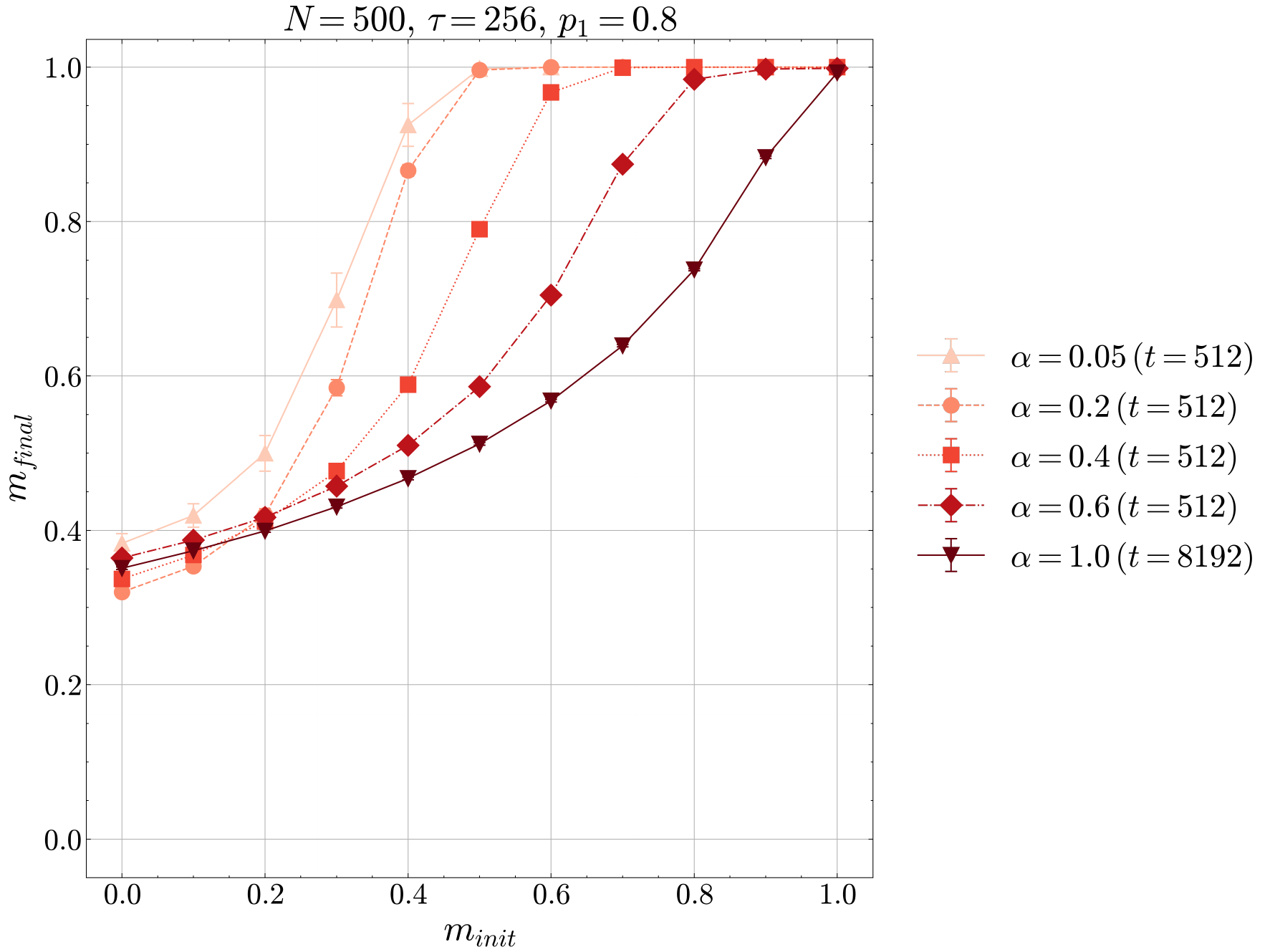}
    \caption{Load dependence of the retrieval maps obtained by centered Daydreaming. The left and right panels show the results for $p_1=0.5$ and $p_1=0.8$, respectively.}
    \label{fig:alpha_dependence}
\end{figure}

We show the retrieval maps in Figure \ref{fig:alpha_dependence}.
In the case with $p_1=0.5$, when $\alpha$ is small, the system reaches the plateau with $m_{\text{final}} \approx 1$ even from small initial magnetizations. 
As $\alpha$ increases, the initial magnetization required for the retrieval to appear becomes larger. 
This behavior reflects the shrinkage of the basin of attraction as the number of stored patterns increases.
For the case with $p_1=0.8$, a similar load dependence is observed. 
For biased patterns, the non-zero value of $m_{\text{final}}$ in the low initial magnetization region includes the overlap induced by the bias. 
Therefore, it cannot be interpreted as a correct retrieval. 
On the other hand, the right panel of Figure \ref{fig:alpha_dependence} shows that, even in the biased case, centered Daydreaming can retrieve stored patterns from finite initial overlaps. 
In particular, for $\alpha<1$, $m_{\text{final}}\approx 1$ is achieved in this retrieval region, indicating that the high-capacity behavior of original Daydreaming for unbiased patterns near $\alpha_{c} \approx 1$ is preserved for biased patterns.
These results support the conclusion that the robustness of centered Daydreaming against bias, as shown in the main text, is not specific to the single load value $\alpha=0.4$.

\section*{Appendix D. Biased Random-Features Hopfield Model}
In the main text, we focused on biased random patterns whose elements are generated independently. 
However, actual memory patterns are not necessarily independent random patterns, and they may have correlations induced by underlying low-dimensional features. 
In this Appendix, we present supplementary experiments that extend the Random-Features Hopfield Model (RFHM) \cite{Negri2023} to the biased case to examine whether centered Daydreaming also works for correlated biased patterns.

In the standard RFHM, structured stored patterns are generated from feature vectors and random coefficients through a linear projection and component-wise $\sign$ nonlinearity.
We append a special feature vector whose components are all equal to $+1$ to introduce a global bias. 
Specifically, we set $f_{i1} = 1$ for all $i$ and $c_{1}^{\nu}= \sqrt{D} \,\mathcal{B}$.
For $k=2, \cdots, D+1$, the feature vectors have i.i.d. uniform binary entries $f_{ik} \in \{+1, -1\}$, and the coefficients $c_k^\nu$ are i.i.d. standard Gaussian variables.
Using these feature vectors and coefficients, the stored patterns are generated as
\begin{align}
    \xi_{i}^{\nu} = 
    \sign \left(\frac{1}{\sqrt{D}} \sum_{k=1}^{D+1} c_{k}^{\nu} f_{ik} \right) = 
    \sign \left(\mathcal{B} + \frac{1}{\sqrt{D}} \sum_{k=2}^{D+1} c_{k}^{\nu} f_{ik} \right).
\end{align}
Here, $\mathcal{B}$ is a parameter that controls the strength of the global bias. 
In particular, when $\mathcal{B}=0$, this model reduces to the standard RFHM.
The contribution from the non-bias features can be written as
\begin{align}
    \eta_{i}^{\nu} = \frac{1}{\sqrt{D}} \sum_{k=2}^{D+1} c_{k}^{\nu} f_{ik} 
    \sim \mathcal{N}(0, 1).
\end{align}
The probability that each component of patterns takes the value $+1$ is given by
\begin{align}
    p_1 = \Pr \left(\xi_{i}^{\nu} = +1 \right) = \Pr \left(\mathcal{B} + \eta_{i}^{\nu} > 0\right).
\end{align}
Therefore, 
\begin{align}
    & p_1 
    = \int_{-\mathcal{B}}^{\infty} \frac{dz}{\sqrt{2 \pi}} \exp \left(-\frac{1}{2} z^2 \right)
    = \frac{1}{2} \left[1 + \erf \left(\frac{\mathcal{B}}{\sqrt{2}} \right)\right],
    \\
    & \mathcal{B}
    =
    \sqrt{2} \, \erf^{-1} (2p_1 - 1).
\end{align}
In the experiments, we determine $\mathcal{B}$ from the above relation for a given $p_1$. 
After learning, we evaluate the retrieval performance by the magnetization with respect to the stored patterns.
We also measure the overlap between the state $\bm{s}$ and the feature vector $\bm{f}$ to examine the feature structure of the RFHM.
\begin{align}
    & m^{\nu} = \frac{1}{N} \sum_{i=1}^{N} \xi_{i}^{\nu} s_{i}
    \\
    & \mu^{k} = \frac{1}{N} \sum_{i=1}^{N} f_{ik} s_{i}
\end{align}
In particular, for the special feature $f_{i1}=1$, the quantity $\mu^1$ is given by
\begin{align}
    \mu^1
    =
    \frac{1}{N}
    \sum_{i=1}^{N}
    s_i .
\end{align}
We call this quantity global feature magnetization.

In the experiments, we set $N=500$, $\alpha=P/N=0.5$, and $\alpha_D=D/N=0.1$. 
We investigate both the standard RFHM with $p_1=0.5$ and the strongly biased case with $p_1=0.8$. 
The inverse learning rate was fixed at $\tau=256$ as in the main text. 
We performed numerical experiments on 20 instances for each setting.

\begin{figure}
    \centering
    \includegraphics[width=0.49\linewidth]{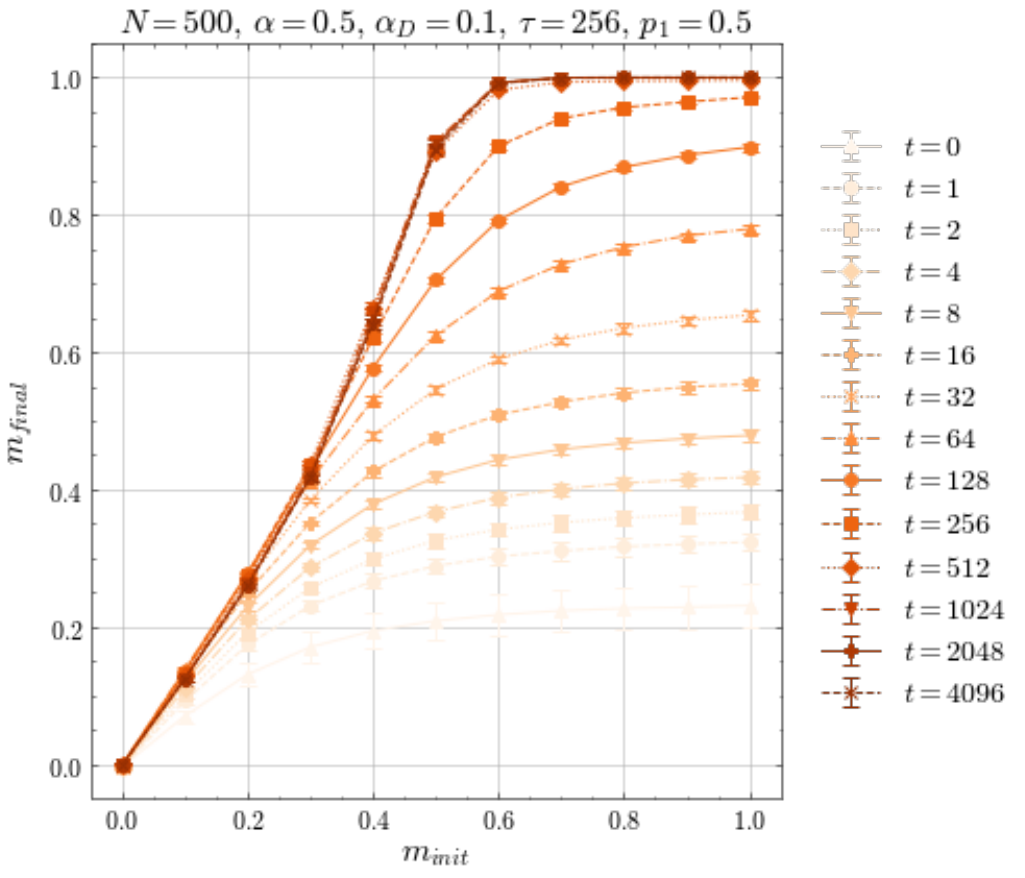}
    \includegraphics[width=0.49\textwidth]{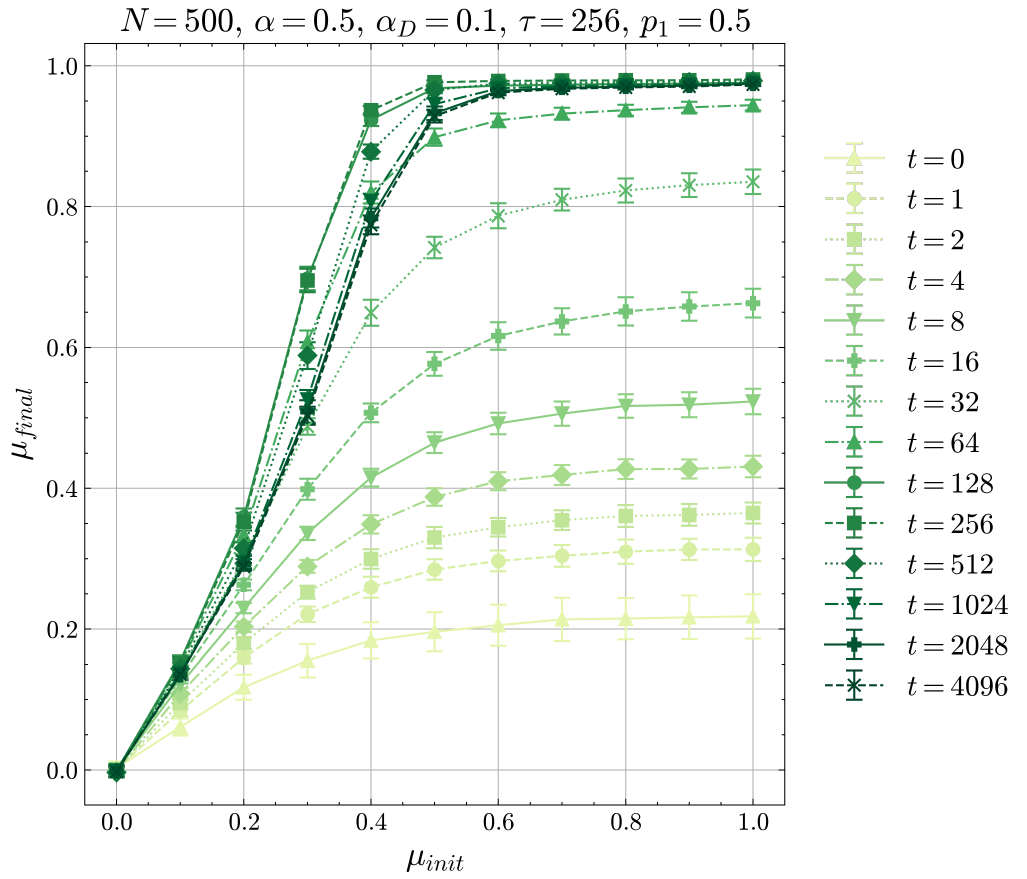}
    \caption{
    Retrieval maps for stored patterns and feature vectors in the standard RFHM with $p_1=0.5$.
    The left and right panels show retrieval maps for the pattern magnetization $m^\nu$ and the non-bias feature magnetization $\mu^k$ with $k\geq 2$, respectively.
    }
    \label{fig:Biased_RFHM_p1_05}
\end{figure}

We first examine the case with $p_1=0.5$. 
In the case of $\mathcal{B}=0$, there is no global bias. 
Therefore, this case corresponds to the conventional RFHM.
We show the results in Figure \ref{fig:Biased_RFHM_p1_05}.
The left panel shows the retrieval map for pattern magnetization $m^\nu$, while the right panel shows the results for the feature magnetization $\mu^k$ with $k \geq 2$. 
After learning, retrieval plateaus with $m_{\text{final}}\approx 1$ and $\mu^k_{\text{final}}\approx 1$ are formed from finite initial overlaps. 
These results are consistent with the previous study \cite{Serricchio2025} on the RFHM and confirm that the experimental setup properly reproduces the standard RFHM.

\begin{figure}
    \centering
    \includegraphics[width=0.5\linewidth]{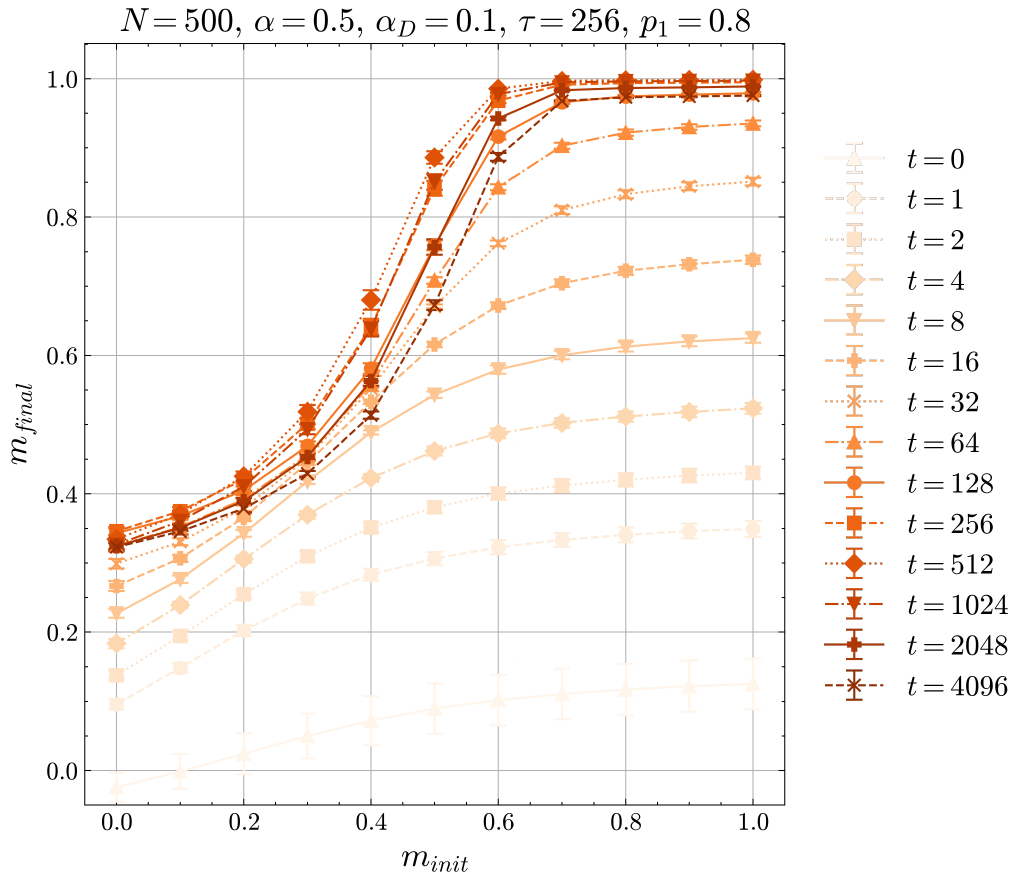}
    \caption{Retrieval map for patterns of the biased RFHM with $p_1=0.8$.}
    \label{fig:Biased_RFHM_p1_08_pattern_magnetization}
\end{figure}

\begin{figure}
    \centering
    \includegraphics[width=0.49\linewidth]{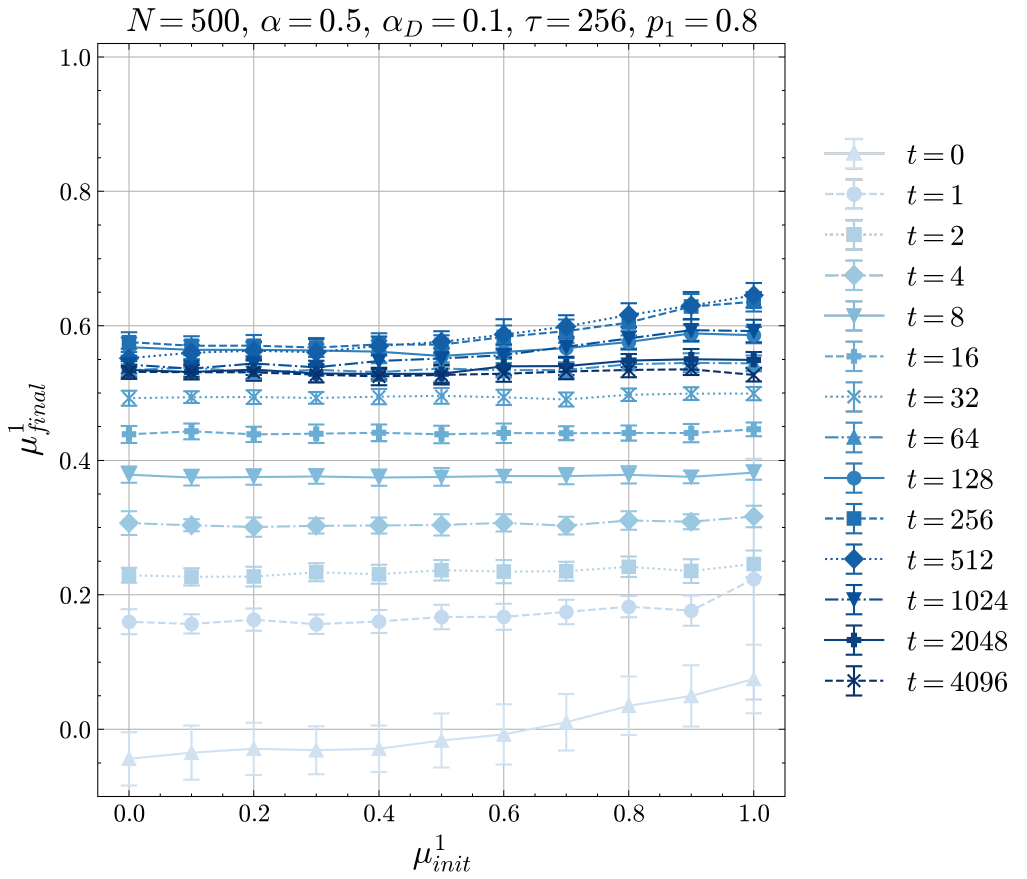}
    \hfill
    \includegraphics[width=0.49\linewidth]{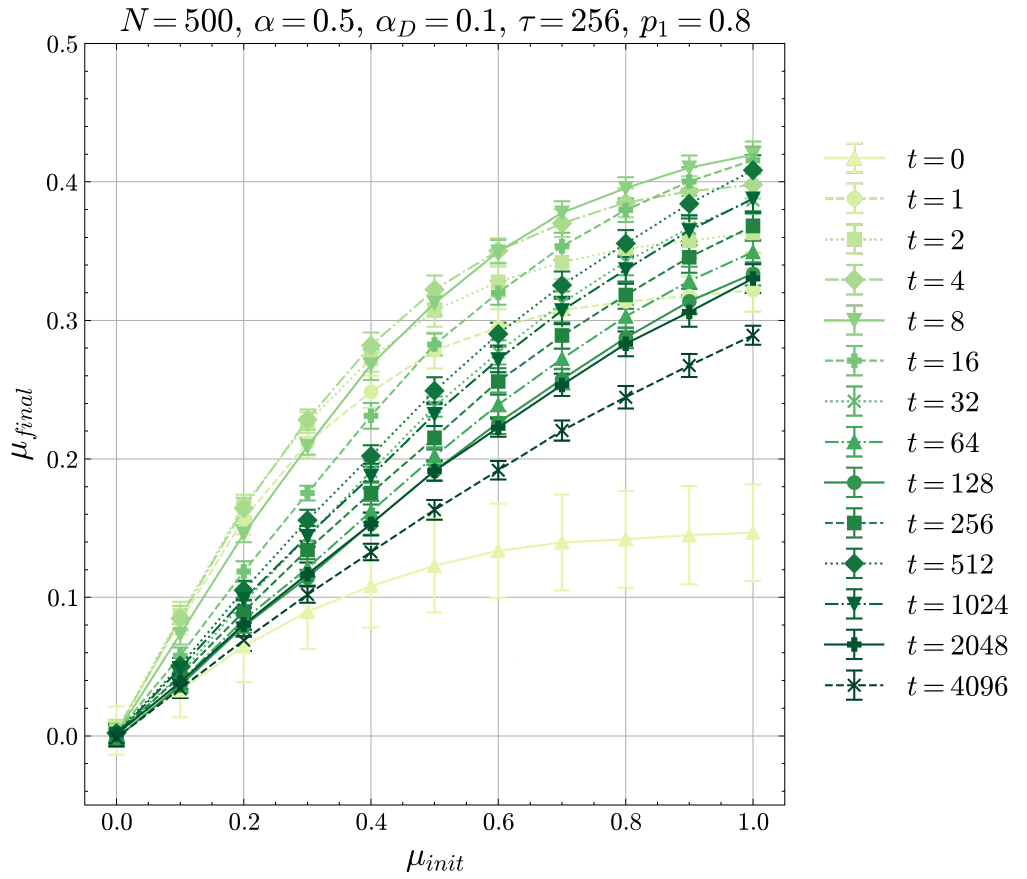}
    \caption{Retrieval maps for feature magnetization of the biased RFHM with $p_1=0.8$.
    The left panel shows the global feature magnetization $\mu^1$, which corresponds to the global magnetization of the state.
    The right panel shows the feature magnetization $\mu^k$ for non-bias features with $k\geq 2$.
    }
    \label{fig:Biased_RFHM_p1_08_feature_magnetization}
\end{figure}

Next, we examine the biased case with $p_1=0.8$. 
In this case, $\mathcal{B}>0$, and the stored patterns contain a strong global bias. 
Therefore, even in the low initial overlap region, a non-zero pattern magnetization appears due to the bias.
Figure \ref{fig:Biased_RFHM_p1_08_pattern_magnetization} shows the results for the pattern magnetization. 
As shown in the figure, even in the presence of a strong global bias, centered Daydreaming forms a retrieval plateau.
However, at $t=2048$ and $t=4096$, we observe a small degradation in the retrieval performance compared with shorter learning epochs. 
This result implies that retrieval performance does not necessarily improve monotonically and converge in the biased RFHM.
Figure \ref{fig:Biased_RFHM_p1_08_feature_magnetization} shows the results for the feature magnetization. 
The left panel shows the global feature magnetization $\mu^1$ corresponding to the special feature $f_{i1}=1$ for all $i$.
The right panel of Figure \ref{fig:Biased_RFHM_p1_08_feature_magnetization} shows the feature magnetization $\mu^k$ for non-bias features with $k\geq 2$. 
As shown in the figure, for $p_1=0.8$, we do not observe either feature magnetization approaching $1$.
These results suggest that, in the biased RFHM, attractors are formed mainly around the stored patterns rather than spontaneously around feature vectors.

Overall, these results confirm that, under a representative setting $(\alpha,\alpha_D)=(0.5,0.1)$, centered Daydreaming can form retrieval plateaus toward stored patterns even for correlated patterns such as the biased RFHM with a global bias. 
These results imply that the effectiveness shown in the main text is not limited to independently generated biased random patterns. 
However, understanding the mechanism behind the slight degradation in retrieval performance after long learning epochs and a more detailed verification of different settings $(\alpha,\alpha_D)$ remain important topics for future work.

\ack{
M.D.\ thanks Dario Bocchi for fruitful discussions.
M.D. and M.O. were funded by programs for bridging the gap between R\&D and IDeal society (Society 5.0) and Generating Economic and social value (BRIDGE) and Cross-ministerial Strategic Innovation Promotion Program (SIP) from the Cabinet Office.
F.R.T.\ thanks Silvio Franz for raising the question that has inspired the present work and the “National Centre for HPC, Big Data and Quantum Computing”, Project CN\_00000013, CUP B83C22002940006, NRRP Mission 4 Component 2 Investment 1.4,  Funded by the European Union - NextGenerationEU, for the financial support.
}

\bibliographystyle{unsrt}
\bibliography{references}

\end{document}